\documentclass[pra,twocolumn,preprintnumbers,amsmath,amssymb,tightenlines,twocolumn,superscriptaddress,epsfig]{revtex4-1}
\usepackage{verbatim}
\usepackage{graphicx}
\usepackage{color}
\usepackage[percent]{overpic}
\usepackage[mathscr]{euscript}

\newcommand{\bra}[1]{\langle\,{#1}\, |}
\newcommand{\ket}[1]{|\,{#1}\,\rangle}
\newcommand{\braket}[2]{\mbox{$\langle\,{#1}\, | \,{#2}\,\rangle$}}

\newcommand{\rvec}[1]{{\mathbf{r}}}

\newcommand{\ssection}[1]{{\noi  \it #1:}}

\newcommand{\sub}[2]{{#1}_{\mbox{\!\! \scriptsize #2}}}

\def\noi{\noindent}
\def\beq{\begin{equation}}
\def\eeq{\end{equation}}

\def\CR{\nonumber\\[0.15cm]}

\newcommand{\rref}[1]{Ref.~\cite{#1}}
\newcommand{\fref}[1]{Fig.~\ref{#1}}

\newcommand{\eref}[1]{Eq.~(\ref{#1})}

\newcommand{\sref}[1]{section~\ref{#1}}

\newcommand{\cref}[1]{chapter~\ref{#1}}

\newcommand{\Cref}[1]{Chapter~\ref{#1}}
\newcommand{\tref}[1]{table~\ref{#1}}

\newcommand{\aref}[1]{appendix~\ref{#1}}
\newcommand{\bref}[1]{(\ref{#1})}

\usepackage{ulem}  
\DeclareMathOperator{\sech}{sech}

\begin{document}

\title{Generation and decoherence of soliton spatial superposition states}
\author{Abhijit Pendse}
\email{apendse@iiserb.ac.in}
\affiliation{Department of Physics, Indian Institute of Science Education and Research, Bhopal, Madhya Pradesh 462 066, India}
\author{Shruti Shirol}
\affiliation{Department of Physics, Indian Institute of Science Education and Research, Bhopal, Madhya Pradesh 462 066, India}
\affiliation{Department of Physics, University of Massachusetts-Amherst, Amherst, MA, 01003, USA}
\author{Shivakant Tiwari}
\affiliation{Department of Physics, Indian Institute of Science Education and Research, Bhopal, Madhya Pradesh 462 066, India}
\author{Sebastian~W\"uster}
\email{sebastian@iiserb.ac.in}
\affiliation{Department of Physics, Indian Institute of Science Education and Research, Bhopal, Madhya Pradesh 462 066, India}
\begin{abstract}
Due to their coherence properties, dilute atomic gas Bose-Einstein condensates seem a versatile platform for controlled creation of mesoscopically entangled states with a large number of particles and also allow controlled studies of their decoherence. However, the creation of such a state intrinsically involves many-body quantum dynamics that cannot be captured by mean-field theory, and thus invalidates the most wide-spread methods for the description of condensates. We follow up on a proposal, in which a condensate cloud as a whole is brought into a superposition of two different spatial locations, by mapping entanglement from a strongly interacting Rydberg atomic system onto the condensate using off-resonant laser dressing [R.~Mukherjee {\it et al.}, Phys.~Rev.~Lett.~{\bf 115} 040401 (2015)]. A variational many-body Ansatz akin to recently developed multi-configurational methods allows us to model this entanglement mapping step explicitly, while still preserving the simplicity of mean-field physics for the description of each branch of the superposition. In the second part of the article, we model the decoherence process due to atom losses in detail. Altogether we confirm earlier estimates, that tightly localized clouds of $400$ atoms can be brought into a quantum superposition of two locations about $3$ $\mu$m apart and remain coherent for about $1$ ms.
\end{abstract}

\maketitle

\section{Introduction} 
%
Ever since the formulation of quantum mechanics in the $1920$s, the quantum to classical transition has been the subject of intense study \cite{schloss_decoherence,quant_class_joos,quant_class_zurek,quant_class_exp_zavatta,quant_class_modi,quant_class_haroche}. One apparent difference between the quantum and classical realms is the existence of quantum coherent superposition states in the former. Decoherence can explain within the usual framework of quantum mechanics why these are typically not observed for macroscopic systems, while the root cause for observing a definite measurement outcome is still not satisfactorily explained within the theory \cite{schloss_decoherence}. This motivates the formulation of collapse models \cite{collapse_bassi_rmp, collapse_romero}, that explore if additional physical laws cause apparently different behaviour of quantum and classical objects.

The unsatisfactory understanding of the quantum to classical transition motivates an experimental drive to bring ever larger controllable quantum systems into superposition states, to check whether they adhere to standard decoherence theory, and hence to the usual framework of quantum mechanics, or whether new physics comes into play. While the creation of truly macroscopic quantum systems remains elusive \cite{macroscopic_quantum_review}, mesoscopic settings that are pushing towards this frontier include matter wave interference in $C_{60}$ molecules \cite{arndt_c60}, organic molecules \cite{organic_gerlich,mlibrary_arndt} or superposition of currents in superconductors \cite{vanderwal_superconducting}. 
There are further proposals to enlarge this pool of candidate platforms for the exploration of the quantum classical boundary through the generation of mesoscopically entangled states in cavity optomechanical systems \cite{opto_dustin}, photon fields in a Kerr medium \cite{macsup_tara}, Rydberg dressed atom clouds \cite{moebius:cat} and in Bose-Einstein Condensates (BECs) \cite{soliton_aparna,barrier_scattering_superposition,sebnrick,bec_superpos_cirac, bec_superpos_pezze,gordon:BECcat,weiss:solitoncat}. 

Here we explore a scheme to generate quantum superposition state of a gaseous BEC that was proposed in \rref{sebnrick} in more detail, where a mesoscopic superposition state of a matter wave bright soliton \cite{aref1,aref2,aref3,aref4,aref5,aref6,aref7,aref8,aref9,aref10,aref11,aref12,aref13,aref14,aref15,aref16} would be prepared by first entangling two control atoms exploiting the Rydberg blockade \cite{gallagher_Rydberg_book,gallagher_Rydberg_review,gaetan:twoatomblock,rblockade_exp_urban}, and then mapping this entanglement onto the bright soliton using Rydberg dressed long-range interactions \cite{sebnrick, dressing_jau, dressing_balewski,dressing_johnson, dressing_wuester, dressing_henkel, dressing_santos, dressing_maucher, dressing_honer, dressing_pupillo}. Initially, a single stationary bright soliton forms a quasi-one dimensional (1D) BEC flanked by two control atoms trapped on either side of it. One then attempts to optically excite these control atoms into a Rydberg state. Since the interaction blockade
prohibits simultaneous excitation of both control atoms, this generates an entangled Bell state where either the one or the other control atom has been excited. Only at this point is the entire soliton subjected to Rydberg dressing, such that it is accelerated away from the excited control atom. Since the system was in a superposition state of either control atom excited, the soliton will evolve into a superposition state of different velocities and later positions.

The first aspect of the above scheme that we describe here in more detail, is the transfer of entanglement from the control atoms onto the mesoscopic BEC bright soliton. Once the latter is in a genuine entangled state, it can no longer be dealt with using mean field theory in which all the condensate particles macroscopically occupy the same single particle state. To nonetheless model the mapping step, we employ a variational many-body Ansatz. For that we assume that the total state of control atoms and Bose-gas may be a superposition, in which each state of the control atom system is entangled with a separate highly occupied orbital for the Bose gas. This scheme goes beyond the standard mean field Gross-Pitaevskii (GP) dynamics by allowing entanglement arising from the interaction of the BEC soliton with the control atoms. Our approach has been guided by recent developments of the Multi-Layer Multi-configurational Time Dependent Hartree for Bosons (ML-MCTDHB) \cite{mlmctdhb_kronke2013, mlmctdhb_pra,mlmctdhb_jcp, mlmctdhb_ion_impurity} used to describe the dynamics of composite systems, each with multiple states. 
Compared to this technique, we keep the Ansatz used here as simple as possible to describe the physics of interest to us. A similar approach was used e.g.~in \cite{Ebgha_compoundatomion_PhysRevA}. 

The second aspect of the experiment proposed in \rref{sebnrick} that we expand upon here, is decoherence of the mesoscopic superposition state after it is created. We expect the primary source of decoherence in this scenario to be atom loss, which can be classified into one, two and three body losses \cite{k1_burt, k1_k3_zundel,k2_roberts,k3_savage,k3_altin}. Here, we use a Lindblad master equation \cite{loss_sinatra1998} to explicitly model the effect of these loss processes on the entangled state of a BEC soliton, while assuming a simplified two-mode model for its spatial dynamics. As is known, the decay of a single atom from the entangled state would destroy the entanglement, hence the combination of all loss processes will govern the timescale on which the mesoscopic entanglement can be sustained \cite{n-1_huang,n-1_lombardo}.

\par 
An advantage of the present superposition state generation scheme is the degree of control over the mesoscopic entanglement generation process. By choosing the Rydberg state of the control atoms and the Rydberg state to which the BEC is dressed, one can control the duration of the required interaction step as well as the final velocity of the soliton.
Simultaneously, also the range of Rydberg-Rydberg interactions and thus the spatial extent of the superposition state can be adjusted. This sets it apart from other proposals to create mesoscopic superposition states in BECs using the collision between two solitons \cite{soliton_aparna}, scattering of solitons from a barrier \cite{barrier_scattering_superposition} or collision between different condensates \cite{bec_superpos_cirac, bec_superpos_pezze}.

This paper is organized as follows. We start in \sref{mapping} with a review of the entanglement transfer scheme proposed in \rref{sebnrick}. After defining the model and Hamiltonian in \sref{hamiltonian}, we then describe our many body Ansatz for modelling the generation of a superposition state of soliton 
locations and present the resultant equations of motions in \sref{ansatz}, while details of the derivation are deferred to \aref{sec:derivation}. For experimentally relevant parameters,
listed fully in \aref{ap:numbers}, we then present our numerical simulations in \sref{generation}. After \sref{mapping} has thus comprehensively treated the generation of a 
soliton spatial superposition state, we then move to its destruction by decoherence in \sref{decoherence}, using a Lindblad master equation parameters of which are derived in \aref{ap:lossrates}. We conclude the article with a discussion of the possible future directions.

\section{Entanglement Mapping} 
\label{mapping}
%
\begin{figure}[htb]
\includegraphics[width=0.9\columnwidth]{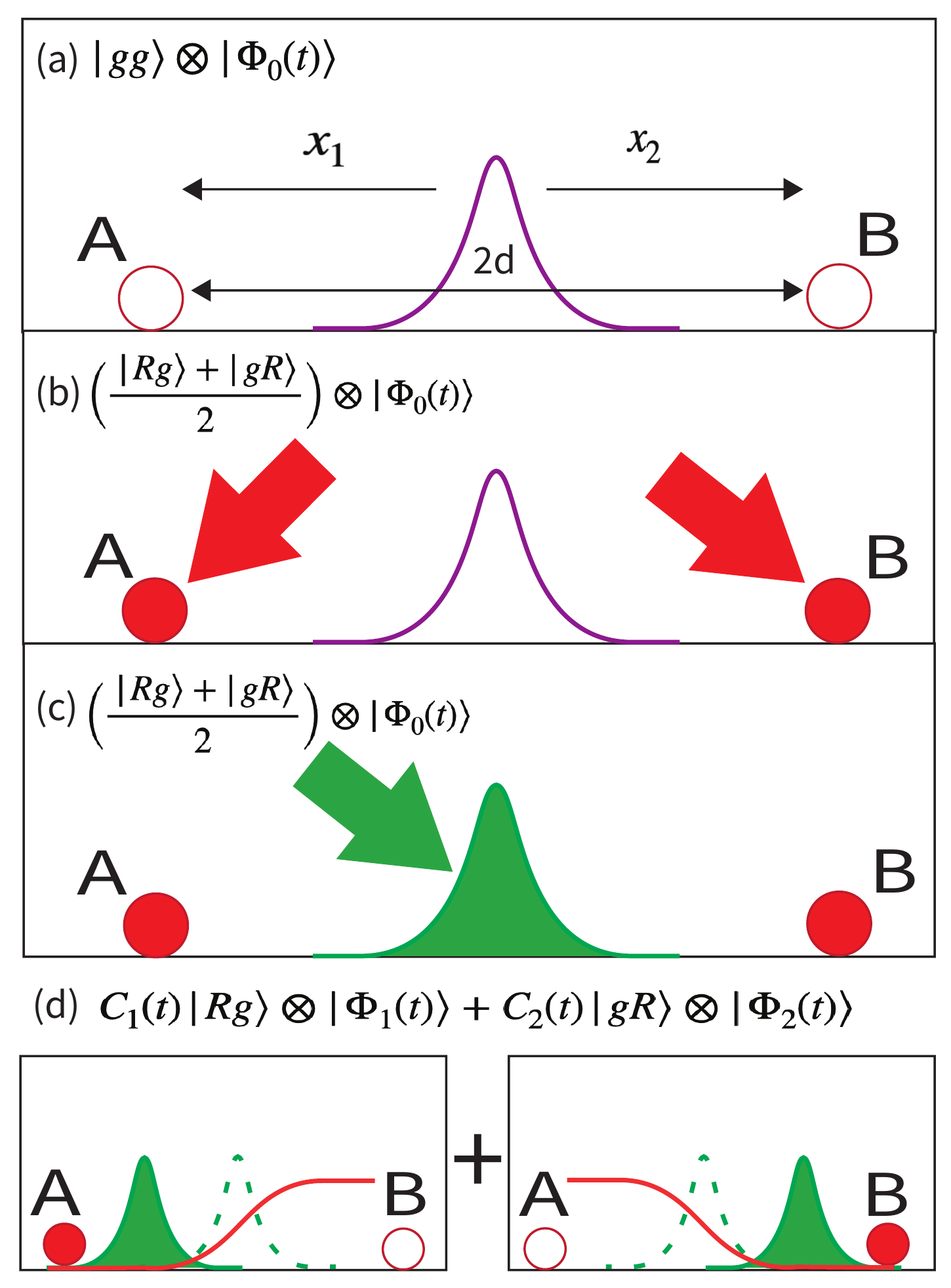}
\caption{Generating a mesoscopically entangled state of a BEC bright soliton state by entanglement transfer from a pair of control atoms, as proposed in \rref{sebnrick}. The purple line sketches the density profile of the soliton and the circles marked $A$ and $B$ represent control atoms. (a) Initially, the soliton is at rest between the control atoms which are both in the ground state $\ket{g}$. The BEC atoms initially are described by the many-body state $\ket{\Phi_{0}}$, with a single macroscopically occupied orbital. (b) A laser (red arrows) targets the control atoms to bring them into a Rydberg excited state $\ket{R}$, but owing to the interaction blockade generates the entangled state shown. (c) We now initiate Rydberg dressing lasers (green), so that all atoms in the soliton (green shade) acquire long range interactions with the control atoms but not among themselves. (d) The resultant long range potential (red line) is conditional on the state of the control atoms, which were in a superposition state. Hence the superposition in electronic states of the control atoms is mapped onto a spatial superposition of the condensate soliton as shown, with $\ket{\Phi_{1}}$ implying the soliton has moved left and  $\ket{\Phi_{2}}$ that it has moved right. Green dashed lines indicate the initial soliton position.}
\label{fig:em_cartoon}
\end{figure}
Let us now briefly review the scheme for the creation of a spatial superposition state of a BEC bright soliton proposed in \rref{sebnrick} in more detail.
Consider an initially stationary bright matter wave soliton consisting of $N$ atoms centered at $x=0$, with two tightly trapped atoms located at a distance $d$, e.g.~$d=1.5$ $\mu$m on either side from the centre of the soliton, as sketched in \fref{fig:em_cartoon} .

We call these tightly trapped atoms ``control atoms". The distance between the control atoms $2d$ is assumed to be less than the blockade radius $R_{b}$ for a particular Rydberg state $\ket{R}=\ket{n_{c},s}$, with principal quantum number $n_c$ and angular momentum $l=0$. We assume that the control atoms can be coupled from their ground state $\ket{g}$ to $\ket{R}$ with Rabi frequency $\Omega_{c}$, e.g.~using a two photon process. This may be done by using laser beams and beam splitters to focus low waist coherent laser source on to the control atoms without affecting the soliton. To enable the control atoms to affect the soliton over the large range $d$, the atoms in the latter are dressed to a Rydberg state $\ket{r}=\ket{n_{d},s}$ with $n_{d}\neq n_{c}$. When attempting to excite the control atoms under blockade conditions, 
they can be brought into the entangled state $\ket{C}=(\ket{gR}+\ket{Rg})/\sqrt{2}$ with high fidelity \cite{gaetan:twoatomblock,rblockade_exp_urban}, where $\ket{gR}$ ($\ket{Rg}$) indicates that the right (left) control atom is excited. It has been shown in \rref{sebnrick} that the Rydberg dressed atoms interact with a Rydberg control atom through an effective potential of the shape
\begin{align}
\sub{U}{eff}(r,t)=V_0(t) \left[1- \left( \frac{r_c}{r}\right)^6 \right]^{-1},
\label{Vint}
\end{align}
 where $r$ is the distance between the impurity atom and the Rydberg dressed atom. The parameters governing the strength $V_0$ and range $r_c$ are given in terms of the underlying Rydberg interactions and dressing parameters in \aref{ap:numbers}. The potential \bref{Vint} is sketched by solid red lines in \fref{fig:em_cartoon}(d), its strength $V_0$ can be controlled in time through the intensity of dressing lasers. 
 
 The control atoms are tightly trapped in their respective positions by an external potential, whereas Bose atoms forming the soliton are un-trapped. Hence, as the latter feel the effective potential $\sub{U}{eff}(r,t)$, they accelerate, setting the soliton into collective motion. Depending upon which control atom is excited, the soliton will either move towards the left or towards the right. The joint state of soliton and control atoms following this conditional acceleration of the soliton then becomes 
\begin{align}
\ket{\Psi(t)}=C_{1}(t) \; \ket{Rg}\otimes\ket{\Phi_{1}(t)}+ C_{2}(t) \; \ket{gR}\otimes\ket{\Phi_{2}(t)}
\end{align}
where $\ket{\Phi_{1,2}(t)}$ are shown in \fref{fig:em_cartoon} and its caption.
After finally de-exciting the control atoms again, we have generated a moving soliton in a superposition state given by $\ket{\Psi}=(\ket{\Phi_{L}}+\ket{\Phi_{R}})/\sqrt{2}$, where $\ket{\Phi_{L}}$, $\ket{\Phi_{R}}$ indicate the left moving soliton state $\ket{\Phi_{1}}$ and right moving soliton state $\ket{\Phi_{2}}$ respectively after they have spatially separated and have negligible spatial overlap. The entire entanglement mapping sequence is sketched in \fref{fig:em_cartoon}.

It is clear that the conversion of the condensate soliton into the superposition state of two different locations $\ket{\Phi_{L}}$ and $\ket{\Phi_{R}}$ inherently cannot be described in the usual mean field picture, since the latter requires all Bosons to occupy the same single particle state. However since creation of the state $\ket{\Psi(t)}$ involves the superposition of just two different types of dynamics, a multi-configurational picture allowing two different highly occupied orbitals, each separately amenable to a mean-field picture, can capture the essentials if each of these orbitals is associated with a specific two-body state of the control atom pair. In the following, we discuss the accordingly customized variational approach for this type of dynamics, which can be classified as a much simplified version of the recently developed  Multi-Layer Multi-configurational Time Dependent Hartree for Bosons (ML-MCTDHB) method \cite{mlmctdhb_kronke2013, Ebgha_compoundatomion_PhysRevA} that can be used to describe the beyond mean-field physics of multi-species Bose gases.

For successfully bringing the soliton into the mesoscopically entangled state, the entire process should be quantum coherent. Hence it is important to assess the strength of decoherence affecting the system. We expect the most important decoherence sources for the soliton to be one-, two- and three body loss of atoms, arising from interactions of condensate atoms with vacuum imperfections or stray photons, spin-changing collisions or inelastic collision between three condensate atoms \cite{loss_sinatra1998}. When considering the soliton as an open quantum system, the loss of an atom can constitute a `{\textit{measurement}}' of the system, leading to a collapse of the wavefunction and breaking the coherence of the mesoscopically entangled state \cite{n-1_huang,n-1_lombardo}. All this will be explored in \sref{decoherence}.

\subsection{Two species model}
\label{hamiltonian}

We first discuss the Hamiltonian of our system, which we split into three parts. One describing the dynamics of the control atoms, $\sub{\hat{H}}{ctrl}(t)$, one for the Bosonic atoms initially constituting the matter wave soliton, $\sub{\hat{H}}{BEC}$, and the last one for interaction of control atoms with the soliton induced by Rydberg dressing, $\hat{H}_{I}(t)$: 
\begin{equation}
\label{eq:hamil1}
\hat{H}=\sub{\hat{H}}{ctrl}(t)+\sub{\hat{H}}{BEC}+\hat{H}_{I}(t).
\end{equation}
As described above, there are two control atoms which can be in a ground-state $\ket{g}$ or Rydberg state $\ket{R}$ under dipole blockade conditions. Thus only the two-body states 
 $\ket{0}\equiv\ket{gg}$, $\ket{1}\equiv\ket{Rg}$ and $\ket{2}\equiv\ket{gR}$ are available to them. We assume the control atoms to be tightly confined in the ground-state of an optical 
 trap, hence no spatial dynamics is allowed for them. Coupling between these electronic states is possible when driving the $\ket{g}\leftrightarrow\ket{R}$ transition, with Hamiltonian
\begin{equation}
\label{eq:hamilcontrol}
\sub{\hat{H}}{ctrl}(t)=\frac{\Omega_{c}(t)}{\sqrt{2}}\Big(\ket{0}\bra{1}+\ket{0}\bra{2}+ \mbox{c.c.}\Big) \otimes \mathbb{I}_{\mathcal{B}},
\end{equation}
where $\Omega_{c}(t)$ is the effective Rabi frequency of that transition and we have included the $\sqrt{2}$ enhancement of the many-body Rabi-frequency \cite{gaetan:twoatomblock}. We denote the Hilbert-space for the control atoms by $\mathcal{C}$, for the Bose atoms by $\mathcal{B}$, such that $ \mathbb{I}_{\mathcal{B}}$ denotes the identity in the space of the Bose atoms.

The Hamiltonian for $N$ Bose atoms each with mass $m$ is given by 
\begin{align}
\label{eq:hamilBEC}
\sub{\hat{H}}{BEC}&=\mathbb{I}_{\mathcal{C}}\otimes\sum_{i=1}^{N}\frac{\hat{p}_{i}^{2}}{2m} +\mathbb{I}_{\mathcal{C}}\otimes\sum_{i=1}^{N}\sub{V}{ext}(\hat{r}_{i})\CR
&+\mathbb{I}_{\mathcal{C}}\otimes\sum_{i,j=1,i\neq j}^{N}\frac{g}{2}\delta(\hat{r}_{i}-\hat{r}_{j}),\nonumber
\end{align}
where $\hat{r}_{i}$ and $\hat{p}_{i}$ are the position and momentum operator respectively of the $i$th atom, $\sub{V}{ext}$ denotes an external potential and $g=4\pi\hbar^{2}a/m$ is the usual contact  interaction strength with s-wave scattering length $a$. The external potential will not be required for the scheme sketched in \fref{fig:em_cartoon}, but shall be included in the derivation of the next section to widen the applicability of the results.

Finally the interaction Hamiltonian is written as
\begin{equation}
\label{eq:HI}
\hat{H}_{I}(t)=\sum_{i=0}^{2}\Big(\ket{i}\bra{i} \otimes \sum_{j=1}^{N}\sub{V}{int}^{(i)}({\bf{x}_{i}}-\hat{r}_{j},t) \Big),
\end{equation}
where for $i\in\{1,2\}$ we have $\sub{V}{int}^{(i)}({\bf{x}_{i}}-\hat{r}_{j},t)=\sub{U}{eff}({\bf{x}_{i}}-\hat{r}_{j},t)$ as the interaction potential between the BEC atoms and the
control atoms at positions $\mathbf{x}_{i}$, see \eref{Vint} and \fref{fig:em_cartoon}. For $i=0$ we define $\sub{V}{int}^{(0)}\equiv0$ for later convenience, since there is no interaction between BEC-atoms and the control atoms when both are in the ground state. The time dependence of the interaction potential is controlled via the intensity of the laser dressing the
BEC atoms off-resonantly to Rydberg states.

\subsection{Variational multi-orbital Ansatz}
\label{ansatz}

The entanglement mapping sequence in \fref{fig:em_cartoon} starts by transferring the control atoms from a product state into an entangled state. This entanglement is then mapped onto a fairly simply structured many-body state for the atoms initially constituting the soliton. To capture this sequence mathematically, we employ the following Ansatz for the many-body wavefunction
 \begin{align}
 \ket{\Psi(t)} &= C_{0}(t) \; \ket{0} \otimes \ket{\Phi_{0}(t)} \; +C_{1}(t) \; \ket{1} \otimes \ket{\Phi_{1}(t)}\CR
 &\hspace{20mm}+C_{2}(t) \; \ket{2} \otimes \ket{\Phi_{2}(t)},
 \label{eq:ansatz}
 \end{align}
 where the $\ket{\Phi_{i}(t)}$ still represent a many body state, namely of the Bose-atoms in space $\mathcal{B}$. The coefficients
$C_{i}(t)$ are the probability amplitudes for each component of the superposition. Now for each $\ket{\Phi_{i}(t)}$, 
we assume the usual mean-field approach for a weakly interacting BEC, and write it in the position space representation as a product  
 \begin{align}
 \braket{\mathbf{r}}{\Phi_{i}(t)}=\Phi_{i}(\mathbf{r},t)=\prod\limits_{j=1}^{N}\phi_{i}(r_{j},t),
  \label{eq:singleorbital}
 \end{align}
where all the $N$ particles occupy the same single particle state $\phi_{i}$. Here $\mathbf{r}$ denotes a vector $\mathbf{r}=[r_1,\dots, r_N]^T$ with all atomic positions except the control atoms. The states $\phi_{i}(r_{j},t)$ are normalized at all times $\int{dr_{j} \; |\phi_{i}(r_{j},t)|^{2}}=1$.

 Using the Hamiltonian in \eref{eq:hamil1} and the Ansatz in \eref{eq:ansatz} with $\Psi(\mathbf{r},t)=\braket{\mathbf{r}}{\Psi(t)}$, we can write the action 
 \begin{equation}
 \begin{split}
 S&=\int{dt\;d^N \mathbf{r}} \Bigg\{\Psi^*(\mathbf{r},t)\Big(\hat{H}-i\hbar\frac{\partial}{\partial t}\Big)\Psi(\mathbf{r},t)\Bigg\}\\
 &-\int{dt\sum_{j=0}^{2}\lambda_{j}\Big(\int{d^N\mathbf{r} \;\Phi^{*}_{j}(\mathbf{r},t)\Phi_{j}(\mathbf{r},t)}-1 \Big)},
 \end{split}
 \end{equation}
where the $\lambda_{j}$ are Lagrange multipliers ensuring the normalization of the many body soliton wavefunction $\Phi_{j}(\mathbf{r},t)$. Importantly, the $\Phi_{j}$ are \emph{not} required to be orthogonal. From the minimization of this action with respect to the coefficients $\{C_{i}(t)\}$ and single particle wavefunctions $\{\phi_{i}(r,t)\}$ in \eref{eq:ansatz}, and exploiting the product forms in \eref{eq:singleorbital}, we obtain our evolution equations as discussed in detail in \aref{sec:derivation}.

The one describing the coefficients $C_{i}(t)$ becomes
\begin{align}
\label{eq:coeff}
i\hbar\frac{\partial}{\partial t}C_{i}(t)&=\sum_{j=0,i\neq j}^{2} C_{j}(t)\bra{i}\sub{\hat{H}}{ctrl}(t)\ket{j} \mathcal{M}_{ij}(t)^{N}\CR
&+N C_{i}(t) \int d\tilde{r} \Big[ \big( \sub{V}{ext}(\tilde{r}) +  \sub{V}{int}^{(i)}({\bf{x_{i}}}-\tilde{r},t)\big) |\phi_{i}(\tilde{r},t)|^2 \CR
&\hspace{25mm}+\frac{g(N-1)}{2}|\phi_{i}(\tilde{r},t)|^{4}\Big],
\end{align}
where $\mathcal{M}_{ij}(t)=\int d\tilde{r} \phi^{*}_{i}(\tilde{r},t)\phi_{j}(\tilde{r},t)$ is the overlap integral between single-particle modes $i$ and $j$. If $\mathcal{M}_{ij}(t)<1$, the orbital attached to state $\ket{i}$ cannot correctly represent the spatial wavefunction of condensate atoms earlier residing in $\ket{\Phi_{j}(t)}$ after those underwent a transition from $\ket{j}$ to $\ket{i}$. If there were a large number of many-body states multiplying $\ket{i}$ in \eref{eq:ansatz} that formed a basis, $\mathcal{M}_{ij}(t)$ would ensure the selection of the appropriate spatial structure. Since this is not the case,  $\mathcal{M}_{ij}(t)<1$ while $\bra{i}\sub{\hat{H}}{ctrl}(t)\ket{j}\neq 0$ signals a limitation of the present Ansatz. We will restrict ourselves to scenarios where this does not occur. The second line in \eref{eq:coeff} simply describes the potential and interaction energy of the $N$ Bosons in single particle state $\phi_{i}$.

The equation of motion of single-particle orbitals is
\begin{align}
\label{eq:sps}
&i\hbar \frac{\partial}{\partial t}\phi_{i}(r,t)=\CR
&\hspace{0mm}\sum_{j=0,i\neq j}^{2} \Bigg\{\Bigg[\frac{C_{j}(t)}{C_{i}(t)}\bra{i}\sub{\hat{H}}{ctrl}(t)\ket{j} \mathcal{M}_{ij}(t)^{N-1}\Bigg]\CR
&\hspace{20mm}\times\Bigg[\phi_{j}(r,t)-\phi_{i}(r,t)\mathcal{M}_{ij}(t)\Bigg]\Bigg\}\CR
&+\hat{h}_{i}\phi_{i}(r,t)+ V_{int}^{(i)}({\bf{x_{i}}}-r,t) \; \phi_{i}(r,t)\CR
&-\phi_{i}(r,t)\int{d\tilde{r}}\phi^{*}_{i}(\tilde{r},t)\Big[\sub{V}{ext}(\tilde{r})\phi_{i}(\tilde{r},t)\CR
&\hspace{3cm}+g(N-1)|\phi_{i}(\tilde{r},t)|^{2}\phi_{i}(\tilde{r},t)\Big]\CR
&-\phi_{i}(r,t)\int{d\tilde{r}\phi^{*}_{i}(\tilde{r},t) V^{(i)}_{int}({\bf{x_{i}}}-\tilde{r},t) \; \phi_{i}(\tilde{r},t)},
\end{align}	
 where the operator $\hat{h}_{i}$ acting on single particle states is given by 
\begin{equation} \hat{h}_{i}=-\frac{\hbar^{2}}{2m}\nabla_{r}^{2}+\sub{V}{ext}(\tilde{r})+g(N-1)|\phi_{i}(\tilde{r},t)|^{2}.
\end{equation}

The approach adopted above can be viewed as a (much) simplified version of the ML-MCTDHB method \cite{mlmctdhb_kronke2013,mlmctdhb_pra,mlmctdhb_ion_impurity,mlmctdhb_jcp}, which is an advanced method to study the dynamics of multi-species ultracold atomic gases. It contains a component akin to basic MCTDHB \cite{mctdhb_alon2008, mctdhb_lode}, where $N$ atoms in one of the species can be distributed among $M$ orbitals in a dynamically evolving manner. Additionally for multiple species, their respective many-body wave functions are allowed to be in several different product states of these multi-orbital superpositions. We confine ourselves to a case where all $N$ atoms are in the same orbital, but this one may differ depending on the state of the second species. The justification for this will be discussed in detail at the end.

In \eref{eq:coeff} and \eref{eq:sps}, the term on the RHS involving integrals result in the usual complex phase oscillations of the $C_{i}$ due to the action of trapping potential on the soliton, the inter-particle interaction of the BEC atoms and the phase imprinted on the orbital $\ket{\Phi_{i}}$ by the control atoms. However, the more interesting dynamics occurs due to the off diagonal parts represented by summations in the two integrals and are dependent on the coupling between the Rydberg states and the ground state of the control atoms. Without this coupling the coefficients corresponding the orbitals $\ket{\Phi_{1}}$ and $\ket{\Phi_{2}}$ would remain zero and as such the system would continue to remain in the initially occupied orbital $\ket{\Phi_{0}}$.

\subsection{Generation of mesoscopic spatial superposition state}
\label{generation}

We now employ the variational method just discussed to model the procedure in \fref{fig:em_cartoon} of bringing a BEC bright soliton into a mesoscopically entangled state, by mapping Rydberg atomic entanglement onto it. Detailed parameters employed throughout the demonstration are given in \aref{ap:numbers}.

\ssection{Step (a)} The starting point at $t<0$ is a quasi 1D bright soliton of $N=400$ atoms of $^{85}$Rb with attractive interactions
as their scattering length is tuned to $a=-5.33\times 10^{-9} m$ using a Feshbach resonance \cite{aref13}. For a quasi-1D setting, the 
radial confinement must be much stronger than the axial one, hence we neglect the latter. The radial trap frequency $\omega_r$ then still affects the effective 1D interaction strength
$g=2\hbar \omega_{r} a$, while in 3D it would have been $g_{3D}=4\pi\hbar^{2}a/m$.

As discussed before, the control atoms are tightly trapped and placed at a distance of $d=1.5$ $\mu$m on each side of the centre of the bright soliton. The total state of the system given by \eref{eq:ansatz} at the beginning is $\ket{\Psi(t<0)}=\ket{0}\otimes\ket{\Phi_{0}(t<0)}$, where $\ket{\Phi_{0}(t<0)}$ is given by \eref{eq:singleorbital} with 
\begin{align}
\label{eq:solitonshape}
\phi_{0}(r,t < 0)&=\frac{1}{\sqrt{2\xi_{0}}}\sech{(r/\xi_{0})}.
\end{align}
The scale $\xi_{0}=\hbar/(m\omega_{r}aN)=0.4$ $\mu$m is the condensate healing length. 

Thus at this point all $N$ atoms form a BEC bright soliton, which is for now amenable to mean field theory. In terms of our formalism in \sref{ansatz} this initial state is described by $C_0(0)=1$, $C_1(0)=C_2(0)=0$ and $\phi_{i}(r,0)$ for $i\in\{0,1,2 \}$ given by \eref{eq:solitonshape}.

\begin{figure}[htb]
\includegraphics[width=0.9\columnwidth]{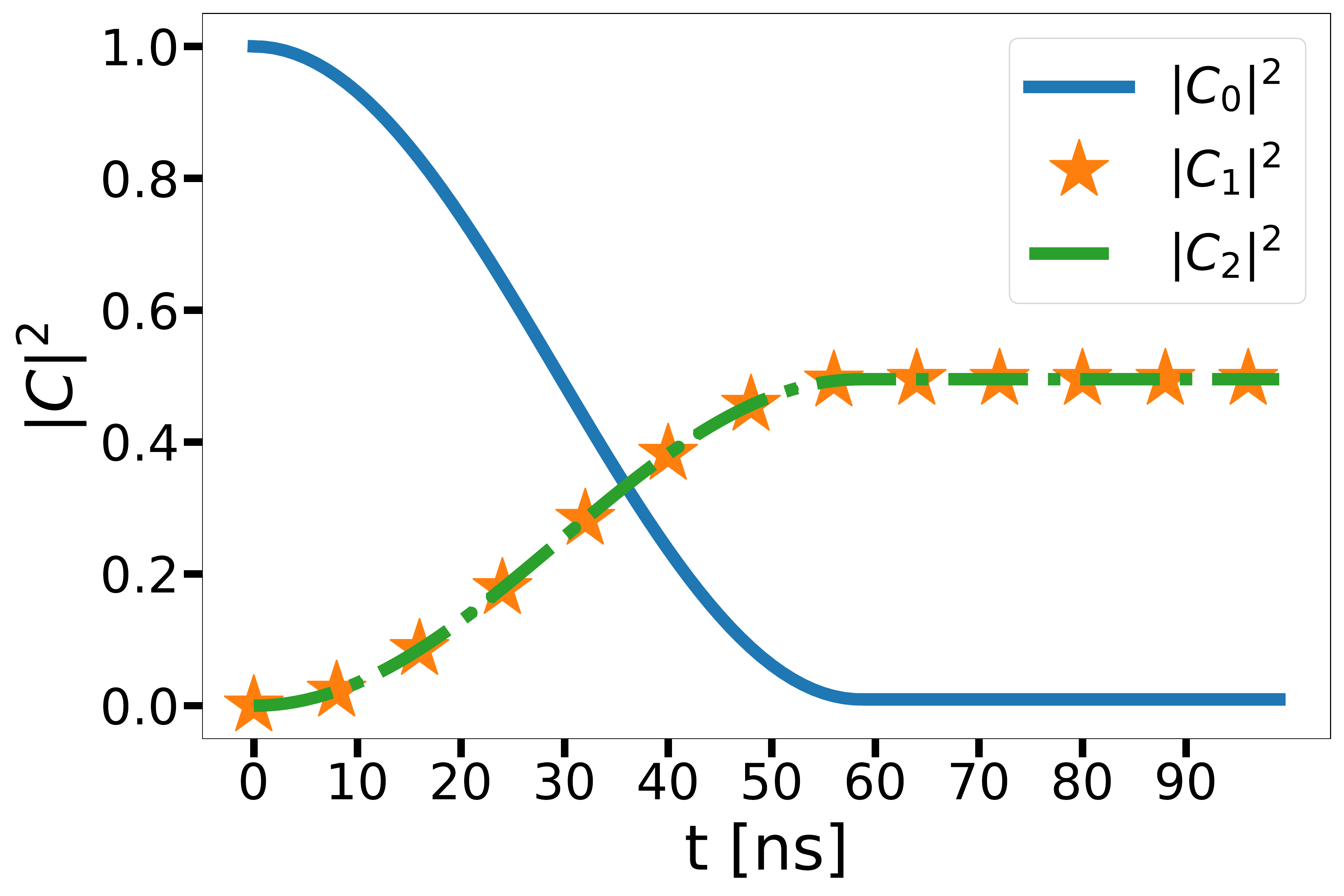}
\caption{Probability of finding the joint system of control atoms and Bose-gas in the many-body states $\ket{0}\otimes\Phi_{0}$ (blue solid line), $\ket{1}\otimes\Phi_{1}$ (yellow star), $\ket{2}\otimes\Phi_{2}$ (green dashed line) according to \eref{eq:coeff} and \eref{eq:sps}, during laser excitation of the control atoms from their ground- to a Rydberg excited state. This corresponds to step (b) in \fref{eq:ansatz} and the description in the text. On this short time scale, the single particle states $\phi_j$ do not evolve significantly.}
	\label{fig:coeff_pop}
\end{figure}
\ssection{Step (b)} The distance $2d$ between the control atoms is within their mutual blockade radius $R_b=(C^{(RR)}_6/\Omega_{c})^{1/6}$, assuming a van-der-Waals interaction with dispersion coefficient $C^{(RR)}$ between two Rydberg atoms in an $\ket{R}$ state. The blockade effectively removes the doubly excited state $\ket{RR}$ from the control atom Hilbert-space, and upon driving Rydberg excitation with Rabi frequency $\Omega_{c}=(3/\sqrt{2\pi^{2}})$ MHz for a short time $\sub{t}{excite}=2.3$ $\mu$s, we can bring the control atoms into the entangled state $(\ket{Rg}+\ket{gR})/\sqrt{2}$. The total system state is hence now $\ket{\Psi(t=\sub{t}{excite})}=(\ket{1}+\ket{2})/\sqrt{2})\otimes\ket{\Phi_{0}(t=\sub{t}{excite})}$. We can explicitly model this step as shown in \fref{fig:coeff_pop}, using \eref{eq:coeff} and \eref{eq:sps}, resulting in coefficients $C_{0}=0$, $C_{1}=C_{2}=1/\sqrt{2}$, while all $\phi_{i}(r,t)$ remain as before since the soliton still is a stationary state of \eref{eq:sps}. Our simulation makes use of the ARK89 \cite{ark,ark_numerical_recipes} adaptive step-size algorithm within the high-level language XMDS \cite{xmds:paper,xmds:citations}.

\ssection{Step (c)} Only at this point does one enable Rydberg dressing of the BEC soliton \cite{sebnrick} for duration $\sub{t}{dress}=36$ $\mu$s, such that $\sub{V}{int}\neq0$ in \eref{eq:HI}. Dressing can be adiabatically enabled and disabled, so that after $\sub{t}{dress}$ all condensate atoms returned to their ground state.
Importantly, since the interaction potential is centered on the particular control atom that is Rydberg excited, the potential $\sub{V}{int}^{(i)}$ entering \eref{eq:sps} is different for $i=1,2$, as sketched in panel (d) of \fref{fig:em_cartoon}, and the potential is absent for $i=0$. This causes the soliton to feel an acceleration into different directions, conditional on the state of the control atom. At the end of this initial acceleration step, at $t=\sub{t}{excite}+\sub{t}{dress}$, the control atoms would ideally be again de-excited into $\ket{0}=\ket{gg}$ by inverting step (b).
We cannot explicitly model that step within the Ansatz in \eref{eq:ansatz} and shall discuss this limitation and a possible remedy later.

\begin{figure}[htb]
\includegraphics[width=0.9\columnwidth]{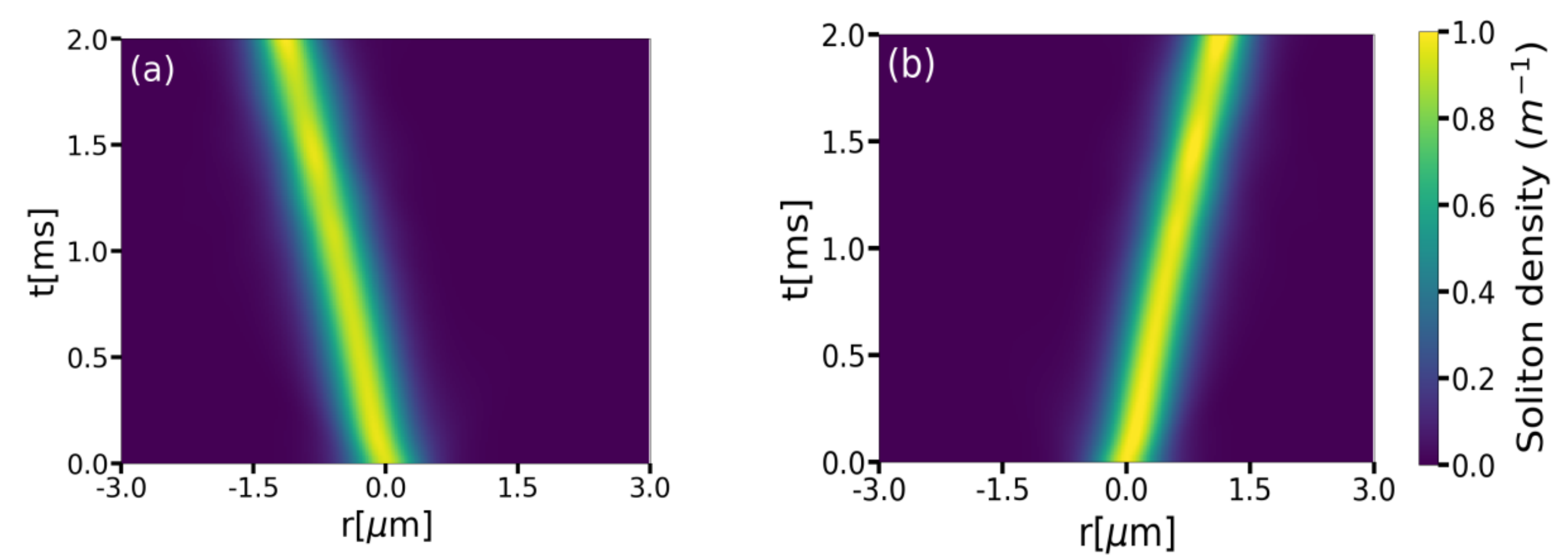}
\caption{Evolution of the individual soliton orbitals (a) $|\phi_{1}|^2$ and (b) $|\phi_{2}|^2$. We can clearly see that owing to the initial acceleration in opposite directions due to interaction with Rydberg control atoms in a superposition state, the soliton finds itself in a superposition of two different modes of motion, through \eref{eq:ansatz}.
\label{fig:solmotion}}
\end{figure}

\ssection{Step (d)} We finally allow free motion of the $N$ Bose gas atoms for a duration $\sub{t}{mov}=2$ ms. As we can see in \fref{fig:solmotion}, in this step the superposition of Rydberg control atoms that we had generated in step (b) is finally converted into a superposition state where the entire soliton of $N$ atoms has either arrived at a location near $\sub{x}{L}\approx-1.5$ $\mu$m, in panel (a) or near $\sub{x}{R}\approx+1.5$ $\mu$m, in panel (b). The two kinds of motion naturally occur in one joint simulation via \eref{eq:coeff} and \eref{eq:sps}, and allow identification of the many-body superposition character through \eref{eq:ansatz}. In terms of \eref{eq:ansatz}, the final state of the simulation at time $\sub{t}{end}=\sub{t}{excite}+\sub{t}{dress}+\sub{t}{mov}$ is now 
 \begin{align}
 \ket{\Psi(\sub{t}{end})} &= C_{1}  \ket{1} \otimes \ket{\Phi_{1}(\sub{t}{end})}+C_{2}(t)  \ket{2} \otimes \ket{\Phi_{2}(\sub{t}{end})},\CR
 &=\frac{1}{\sqrt{2}}\left(\ket{1} \otimes \ket{\Phi_{R}}+ \ket{2} \otimes \ket{\Phi_{L}} \right),
 \label{eq:finalstate}
 \end{align}
where the position space representation of $\ket{\Phi_{L,R}}$ is $\braket{{\bf{r}}}{\Phi_{L,R}}= \prod\limits_{j=1}^{N}\phi_{L,R}(r_{j} )$. Here $\phi_{L,R}$ represent the single particle states $\phi_{1,2}$ when they have significantly separated, such that their overlap $\int{dr}\phi_{1}^{*}(r)\phi_{2}(r) \ll 1$.

 At this point we have thus extended \rref{sebnrick} by an explicit calculation of the entangling many-body dynamics yielding \eref{eq:finalstate}.

Prior to the phase of free motion in step (d), the control atoms should be driven back in to their ground state $\ket{0}$ to disentangle them from the BEC and avoid decoherence through their spontaneous decay. This step cannot be modelled yet with the present Ansatz, since it would require the attachment of at least two orbitals attached to the $\ket{0}$ state of the control atoms. We defer this to future work, since providing the variational Ansatz with further orbitals my cause convergence problems due to centre of mass diffusion of the soliton \cite{mctdhb_convergence_cosme,mctdhb_convergence_weiss, soliton_aparna}. While our restriction to a single orbital per impurity state is of course also not a converged many-body theory, it is still expected to capture the essence of mesoscopically entangled state creation up to the point of control de-excitation.

\section{Decoherence of mesoscopic spatial superposition state}
\label{decoherence}
%
Atoms in the mesoscopic superposition state generated above are not isolated but interact with their environment. Two important components of this interaction are collision with  residual uncondensed $^{85}$Rb atoms, as well as inelastic collisions within the condensate, which result in ejection of atoms from it and hence loss into the environment.
Through these the many-body system discussed so far will decohere, and we expect the mesoscopic superposition state to be fragile. The loss of a single atom from such a state is known to decohere the state \cite{n-1_huang,n-1_lombardo}. A detailed assessment of decoherence times is necessary for many practical uses of mesoscopic superposition states, since any bounds on parameters for collapse models \cite{collapse_bassi_rmp,collapse_romero} can only arise if these affect the state prior to decoherence. 

We can neglect the decoherence induced in the soliton during excitation of control atoms into the superposition state and Rydberg dressing of soliton atoms: The control atoms can be excited within tens of nanoseconds, much shorter than the timescale of loss processes in the BEC. Further, the Rydberg dressing can operate far detuned with parameters listed in \aref{ap:lossrates}, so that the relative fraction of Rydberg state versus ground-state populations is of the order of $1\times 10^{-5}$. This ensures that also spontanous Rydberg decay during the dressing does not significantly affect the BEC on the relevant time scale of $\sub{t}{dress}=36\mu s$ \cite{dressing_loss_aman,dressing_loss_plodzien}.

Since the duration of step (d) in \sref{generation}, in which the soliton moves freely, is orders of magnitude larger than the preceding steps, we conclude that decoherence is most relevant during this phase.
The three major loss mechanisms in a BEC are (i) one body loss due to collision of the condensate atoms with the atoms in the thermal cloud stray photons or vacuum imperfections, (ii) the spin-flipping two body interactions which results in the loss of condensate atoms from the trap and (iii) thirdly the three body losses due to inelastic collisions between the condensate atoms \cite{loss_sinatra1998}. In what follows, we model our system using a Lindblad Master equation and calculate the time required for the loss of an atom from the soliton. If this time is larger than the time scale of the experiments performed, then we can successfully create a macroscopic superposition state of BEC which can be tested in experiments. 

Note that the only contribution of thermal cloud interactions with condensate atoms that we consider here is when these cause an atom to be lost from the condensate.  Another contribution could be decoherence from elastic collisions with thermal cloud atoms. It has been shown in \cite{heating_effect_thesis1, heating_effect_thesis2} that the rate for the latter is however less than the former, hence we have neglected it in our estimate.

%
\subsection{Two-mode model and Decoherence sources}
\label{TMM}
%
To render the treatment of decoherence with a master equation tractable, we use a two mode model describing the atom number dynamics, with a restriction of spatial modes to a ``left" and a ``right" moving soliton mode, that are spatially separated. These would correspond to $\phi_{1}(\mathbf{r},t)$ and $\phi_{2}(\mathbf{r},t)$ created after step (c) in \sref{generation}. We utilize Fock states, where $\ket{0N}$ represents all $N$ atoms residing in the right moving bright soliton mode and $\ket{N0}$ correspondingly all atoms in the left moving one. A decay of atoms from these states will cause incoherent population transfer to $\ket{0M}$ and $\ket{M0}$ with $M<N$. As described above, major loss processes are one-, two- and three-body losses, which affect the density of the condensate $n(t)$ as \cite{roberts_phd}
\begin{equation}
\frac{\partial}{\partial t}n(t)=-\kappa_{1}n(t) -\kappa_{2} n^{2}(t)-\kappa_{3}n^{3}(t),
\label{eq:densityloss}
\end{equation}
where $\kappa_{1}$, $\kappa_{2}$ and $\kappa_{3}$ denote the one, two and three body loss rate coefficients respectively. For $^{85}$Rb we take loss-rate coefficients $\kappa_{1}=6\times 10^{-3} \; s^{-1}$ \cite{k1_burt,k1_k3_zundel} for single-body loss, $\kappa_{2}= 2\times 10^{-20} m^{3}s^{-1}$ \cite{k2_roberts} for two-body loss and $\kappa_{3}=2\times 10^{-40} \; m^{6}s^{-1}$ \cite{k3_altin,k3_savage} for three-body loss. Note, that these loss rates vary significantly between atomic species, isoptopes, spin-states and in the case of one-body loss even experimental setup. Here we have chosen values for each loss-rate that we consider representative.

We now model the decoherence arising from these loss processes with a Lindblad master equation for the density matrix of the system
\begin{align}
\label{eq:densmat}
\hat{\rho}(t)=\sum\limits_{k,l,m,n =0}^{N} \rho_{kl;mn}(t)\ket{kl}\bra{mn}, 
\end{align}
where $\ket{nm}$ are Fock-states in the two-mode model as discussed above. 

Following \cite{loss_sinatra1998, jack}, the starting point is the master equation 
\begin{equation}
\begin{split}
\frac{d\hat{\rho}(t)}{dt}&=\frac{1}{i\hbar}[\hat{H},\hat{\rho}(t)]\\
&+\int{}d{r}\sum_{n=1}^{3}\Big(\kappa_{n}[\hat{\Psi}(r)]^{n}\rho(t)[\hat{\Psi}^{\dagger}(r)]^{n}\\
&\hspace{2cm}-\frac{\kappa_{n}}{2}\Big\{[\hat{\Psi}^{\dagger}(r)]^{n}[\hat{\Psi}(r)]^{n},\rho(t)\Big\}\Big),
\end{split}
\label{eq:lme1}
\end{equation}
where $\hat{\Psi}(r)$/ $\hat{\Psi}^{\dagger}(r)$ are field operators annihilating/ creating a Boson at a position $r$, and the sum over $n$ thus lists the three different loss-processes. The Hamiltonian $\hat{H}$ is as given in \eref{eq:hamil1}. In our simple model with two spatial modes, field operators become
\begin{equation}
\label{eq:lme2}
\hat{\Psi}(r)=\phi_{L}(\mathbf{r})\hat{a}+\phi_{R}(\mathbf{r})\hat{b},
\end{equation}
where $\phi_{L}(\mathbf{r})$ and $\phi_{R}(\mathbf{r})$ represent the single particle states for the sequence of entanglement mapping discussed in the previous section when the modes have separated. In other words, $\phi_{L}(\mathbf{r})=\phi_{1}(\mathbf{r})$ and $\phi_{R}(\mathbf{r})=\phi_{2}(\mathbf{r})$ at a time where $\phi_{1,2}$ have reached negligible spatial overlap. Here $\hat{a}$ and $\hat{b}$ represent the annihilation operator for the left and the right mode respectively. Insertion into \eref{eq:lme1} and projection onto $\ket{kl}\bra{mn}$ 
yields the evolution equation for the coefficients of the density matrix $\rho(t)_{kl;mn}$
\begin{equation}
\label{eq:decoherence}
\begin{split}
\frac{\partial \rho_{kl;mn}(t)}{\partial t}&= \mathcal{T}_{0;klmn} \; \rho_{kl;mn}(t) \\
&\hspace{-1.5cm}+\bar{\kappa}_{1}\Big(\mathcal{T}_{1;km} \; \rho_{(k+1)l;(m+1)n}(t)+\mathcal{T}_{1;ln} \; \rho_{k(l+1);m(n+1)}(t)\Big)\\
&\hspace{-1.5cm}+\bar{\kappa}_{2}\Big(\mathcal{T}_{2;km} \; \rho_{(k+2)l;(m+2)n}(t)+\mathcal{T}_{2;ln} \; \rho_{k(l+2);m(n+2)}(t)\Big)\\
&\hspace{-1.5cm}+\bar{\kappa}_{3}\Big(\mathcal{T}_{3;km} \; \rho_{(k+3)l;(m+3)n}(t)+\mathcal{T}_{3;ln} \; \rho_{k(l+3);m(n+3)}(t)\Big).
\end{split}
\end{equation}
The details of the derivation, along with the expression for the effective loss rate coefficients $\bar{\kappa}_{j}$ and combinatorial factors $\mathcal{T}$s are presented in \aref{ap:lossrates}. The $\bar{\kappa}_{j}$ are based on the $\kappa_{j}$ defined at the beginning of this section, but then also are sensitive to the soliton mode shape $\phi_{L,R}(\mathbf{r})$.

To significantly simplify the equation above, we have assumed that the overlap of the two modes vanishes $\int{d^{N}\mathbf{r} \; \phi^{*}_{L}(\mathbf{r})\phi_{R}(\mathbf{r})}=0$, which strictly means that the calculation is accurate only after the soliton in different branches of the superposition has moved by its width. In \fref{fig:solmotion} it has done so after approximately $0.5$ ms, which makes this a good approximation for three-quarters of the relevant evolution period.

\begin{figure}[htb]
\includegraphics[width=0.9\columnwidth]{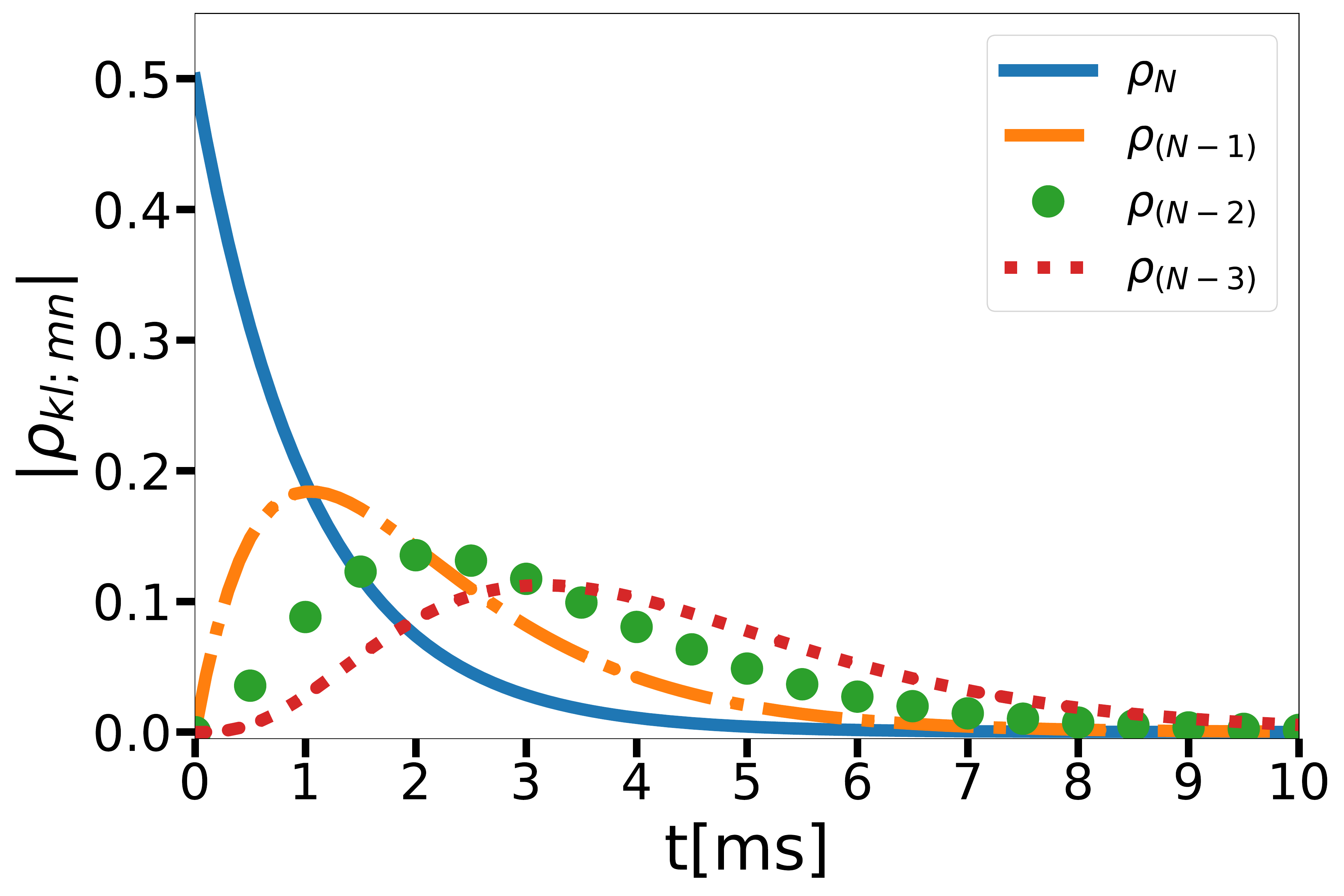}
	\caption{ Evolution of density matrix components in the presence of atom losses, starting from a mesoscopically entangled state with $N=400$ atoms. Figure shows the modulus of selected components as a function of time. The (solid blue line) is $\rho_{N}$, corresponding to populations $\rho_{0N;0N}$, $\rho_{N0;N0}$ and coherences $\rho_{0N;N0}$ and $\rho_{N0;0N}$, which all have identical time evolution. We also show $\rho_{N-1}$ (dash-dot orange line), $\rho_{N-2}$ (green circles) and $\rho_{N-3}$ (dotted red line), where the elements $\rho_{i}$ represent populations $\rho_{0i;0i}$ and $\rho_{i0;i0}$ for all $i<N$, the corresponding coherences $\rho_{i0;0i}$ and $\rho_{0i;i0}$ for $i<N$ remain zero throughout.}
	\label{fig:den_mat_ns}
\end{figure}	
In the two mode model, the mesoscopically entangled state of \sref{generation} is represented by the
 initial state $\ket{\Psi(t=0)}=(\ket{N0}+\ket{0N})/\sqrt{2}$, which gives a density matrix $\hat{\rho}(t=0)=\frac{1}{2}(\ket{N0}\bra{N0}+\ket{N0}\bra{0N}+\ket{0N}\bra{N0}+\ket{0N}\bra{0N})$. Thus the only initially non-vanishing elements of the density matrix are $\rho_{N0;N0}=\rho_{N0;0N}=\rho_{0N;N0}=\rho_{0N;0N}=1/2$.
This corresponds to the final result of \sref{generation}, if we have de-excited the control atoms. The \eref{eq:decoherence} can in principle be solved analytically, using the thermo-field technique \cite{Chaturvedi_thermofield}, however here we resort to a numerical solution which is shown in \fref{fig:den_mat_ns} for $N=400$. 

We show several selected density matrix elements in \fref{fig:den_mat_ns}. All those discussed above, which are initially nonzero, follow the same time evolution shown as solid blue line. 
Except $\rho_{0N;N0}$ and $\rho_{N0;0N}$, no coherence matrix elements become populated, hence the density matrix no longer significantly contains coherences after roughly $3$ ms. Instead we see a rise of population matrix elements for fewer atoms. For even later times density matrix populations with even smaller atom content becomes populated, which we do not show.

\begin{figure}[h]
	\includegraphics[width=0.9\columnwidth]{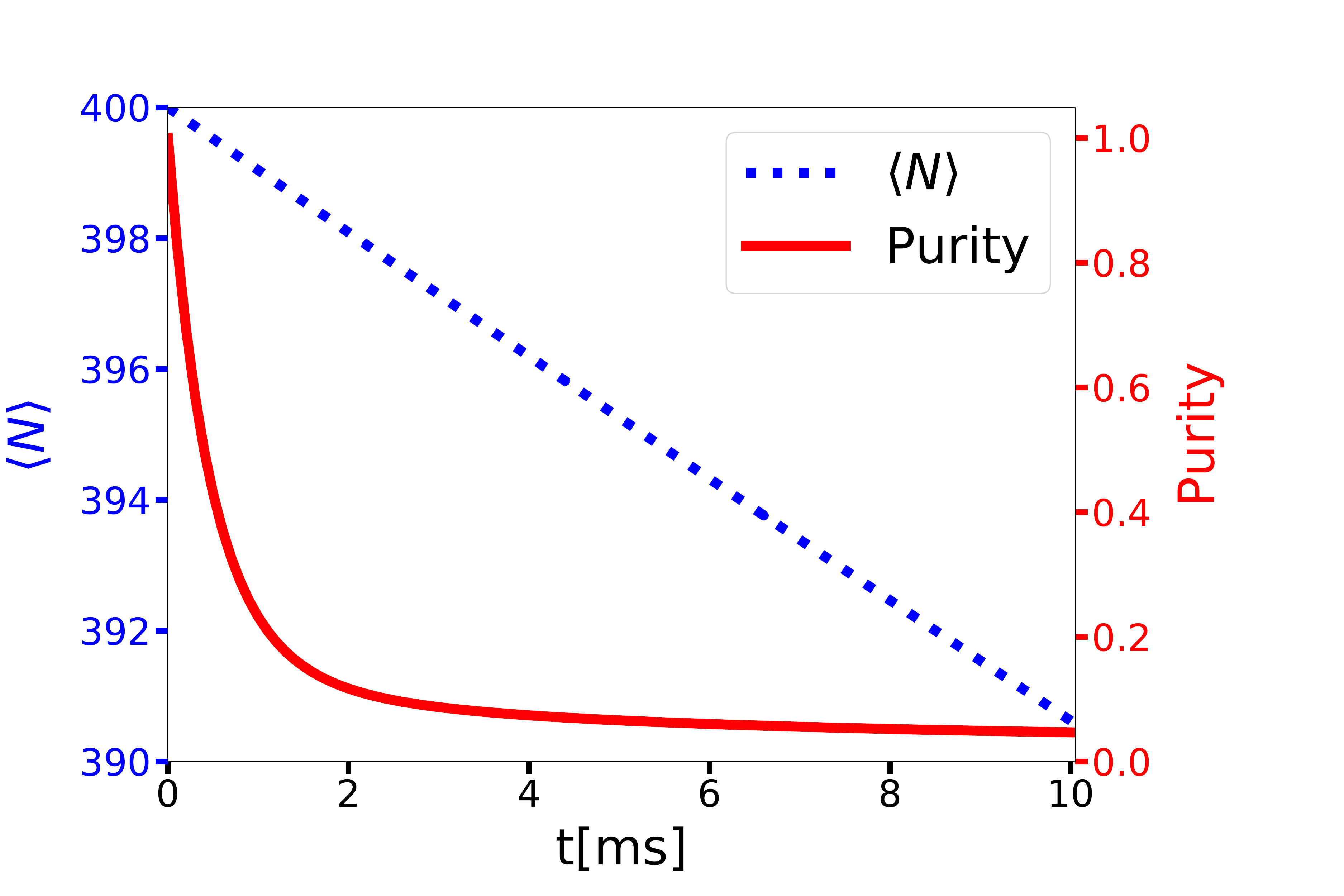}
	\caption{Evolution of the average number of atoms $\langle \hat{N} \rangle $ from \eref{eq:meanN} (dotted blue line) and purity from \eref{eq:purity} (solid red line) of the system during the initial $10$ ms. After a single atom is lost on average, around time $1$ ms, the purity indicates almost complete decoherence to a mixed state. }
	\label{fig:nexp_purity_combo}
\end{figure}

To display the ramifications of this more clearly, we further calculate the average number of particles 
\begin{equation}
\label{eq:meanN}
\langle \hat{N} \rangle =\sum_{k,l=0}^{N}\rho_{kl;kl}(t) (k+l)
\end{equation}
and the purity of the density matrix as a function of time,
\begin{equation}
\label{eq:purity}
P=\mbox{Tr}[\hat{\rho}^2]=\sum\limits_{k,l,m,n =0}^{N}|\rho_{kl;mn}(t)|^{2},
\end{equation}
shown in \fref{fig:nexp_purity_combo}. We see that by the time the system has lost just one atom on average, the initially complete purity has almost entirely disappeared.
After rapid initial decoherence, the purity decay slows significantly but it still continues to drop on much larger time-scales. 
It is bounded from below by the dimension of the computational Hilbertspace, which is $d=N\times N$, but this minimum is not approached yet for the times shown in \fref{fig:nexp_purity_combo}.
That the mesoscopically entangled state is decohered essentially at the time a single atom is lost is in accordance with earlier studies pertaining to BEC \cite{cat_brune,bec_superpos_cirac, cat_pawlowski, cat_zurek}. It has been discussed in  \rref{cat_savage} that the coherence of a mesoscopically entangled state decays exponentially, with a time scale that will in our case be proportional to the number of atoms. Thus the larger the number of atoms, the faster is the decoherence. This can be seen directly from \eref{eq:decoherence} and the constant factors involved therein. As mentioned previously, for the parameters of our system the most prominent loss is the one-body loss. Therefore looking at Eqs.(\ref{eq:rate_t0}) and (\ref{eq:rate_t1}), one can see that the loss coefficient contains the number of atoms $N$ explicitly as well as implicitly in the factors $k,l,m,n$. One can then conclude from these equations that increasing the number of atoms in the system will accelerate the decay processes and consequently accelerate the fall of purity as well as the average atom number. 

We have re-produced this phenomenology here, for the specific parameters of the sequence in \fref{fig:em_cartoon}, in order to be able to accurately predict the expected decoherence time-scale $\tau\approx 1$ ms. The results here confirm the estimates given in \rref{sebnrick}. By comparing simulations with different loss modes separately, we further identified single body loss as chief decoherence mechanism for the scenario here, using parameters listed above. 

\section{Conclusions and outlook} 

We have studied the proposal of entanglement transfer from two Rydberg control atoms in the blockade regime onto a BEC bright soliton of \rref{sebnrick} in more detail. Since \cite{sebnrick} did focus on engineering the interactions \bref{Vint} between dressed ground-state atoms and a Rydberg control atom, the proposal to create a mesoscopically entangled state relied on physical arguments without explicit simulation, and decoherence time-scales were estimated. Here we have provided a formal theoretical framework for both these aspects, going beyond mean field theory and considering the underlying many-body problem.

We confirm, that within a creation time of about $40$ $\mu$s, a BEC bright soliton containing about $400$ atoms can be quantum entangled with two control atoms. Conditional on which of the two control atoms is Rydberg excited, the entanglement corresponds to the soliton having received a momentum kick in either of two opposite direction. After a further evolution time of about $1$ ms, this  momentum kick can be converted into a significant distance travelled, so that the solitons finally find themselves in a superposition of two locations about $3$ $\mu$m apart. 

Our model of this process is based on a restricted multi-configurational wave-function with just three orbitals, tied to three control atom basis states. This is motivated by simplicity and physically by the structure of the interaction Hamiltonian between the BEC and the control atoms, which will to leading order create a state of this form. In principle the Ansatz can be augmented to  a larger number of superposition components, more orbitals per components and varying atom numbers per orbital, leading ultimately to the full fledged multi-component MCTDHB form.

After the overlaping orbitals have separated, they undergo free motion.  During that relatively long final phase of free motion, the mesoscopically entangled state may suffer decoherence, which we have explored comprehensively using a Lindblad Master equation. We confirm the earlier estimate of decoherence time of about a few milliseconds.

The variational formalism discussed here could be useful also for other scenarios where entanglement between an initially pure Bose-Einstein condensed cloud and Rydberg impurity atoms is generated, for example if a Rydberg impurity in a quantum superposition is embedded directly into the BEC \cite{sid_preparation}, or if a multi-atom Rydberg crystal \cite{pohl:crystal, rickvB:adiab_crystals, rcrystal_schauss1, rcrystal_schauss2}, is generated within a BEC and then diagnosed through its interaction with it. 
 
\acknowledgments
We thank Thomas Busch and Subhash Chaturvedi for fruitful discussions and the Max-Planck society for financial support under the MPG-IISER partner group program. 
\bibliography{cat_coherence}

\begin{thebibliography}{91}%
\makeatletter
\providecommand \@ifxundefined [1]{%
 \@ifx{#1\undefined}
}%
\providecommand \@ifnum [1]{%
 \ifnum #1\expandafter \@firstoftwo
 \else \expandafter \@secondoftwo
 \fi
}%
\providecommand \@ifx [1]{%
 \ifx #1\expandafter \@firstoftwo
 \else \expandafter \@secondoftwo
 \fi
}%
\providecommand \natexlab [1]{#1}%
\providecommand \enquote  [1]{``#1''}%
\providecommand \bibnamefont  [1]{#1}%
\providecommand \bibfnamefont [1]{#1}%
\providecommand \citenamefont [1]{#1}%
\providecommand \href@noop [0]{\@secondoftwo}%
\providecommand \href [0]{\begingroup \@sanitize@url \@href}%
\providecommand \@href[1]{\@@startlink{#1}\@@href}%
\providecommand \@@href[1]{\endgroup#1\@@endlink}%
\providecommand \@sanitize@url [0]{\catcode `\\12\catcode `\$12\catcode
  `\&12\catcode `\#12\catcode `\^12\catcode `\_12\catcode `\%12\relax}%
\providecommand \@@startlink[1]{}%
\providecommand \@@endlink[0]{}%
\providecommand \url  [0]{\begingroup\@sanitize@url \@url }%
\providecommand \@url [1]{\endgroup\@href {#1}{\urlprefix }}%
\providecommand \urlprefix  [0]{URL }%
\providecommand \Eprint [0]{\href }%
\providecommand \doibase [0]{http://dx.doi.org/}%
\providecommand \selectlanguage [0]{\@gobble}%
\providecommand \bibinfo  [0]{\@secondoftwo}%
\providecommand \bibfield  [0]{\@secondoftwo}%
\providecommand \translation [1]{[#1]}%
\providecommand \BibitemOpen [0]{}%
\providecommand \bibitemStop [0]{}%
\providecommand \bibitemNoStop [0]{.\EOS\space}%
\providecommand \EOS [0]{\spacefactor3000\relax}%
\providecommand \BibitemShut  [1]{\csname bibitem#1\endcsname}%
\let\auto@bib@innerbib\@empty
\bibitem [{\citenamefont {Schlosshauer}(2007)}]{schloss_decoherence}%
  \BibitemOpen
  \bibfield  {author} {\bibinfo {author} {\bibfnamefont {M.~A.}\ \bibnamefont
  {Schlosshauer}},\ }\href@noop {} {\emph {\bibinfo {title} {Decoherence: and
  the quantum-to-classical transition}}}\ (\bibinfo  {publisher} {Springer
  Science \& Business Media},\ \bibinfo {year} {2007})\BibitemShut {NoStop}%
\bibitem [{\citenamefont {Joos}\ \emph {et~al.}(2013)\citenamefont {Joos},
  \citenamefont {Zeh}, \citenamefont {Kiefer}, \citenamefont {Giulini},
  \citenamefont {Kupsch},\ and\ \citenamefont {Stamatescu}}]{quant_class_joos}%
  \BibitemOpen
  \bibfield  {author} {\bibinfo {author} {\bibfnamefont {E.}~\bibnamefont
  {Joos}}, \bibinfo {author} {\bibfnamefont {H.~D.}\ \bibnamefont {Zeh}},
  \bibinfo {author} {\bibfnamefont {C.}~\bibnamefont {Kiefer}}, \bibinfo
  {author} {\bibfnamefont {D.~J.}\ \bibnamefont {Giulini}}, \bibinfo {author}
  {\bibfnamefont {J.}~\bibnamefont {Kupsch}}, \ and\ \bibinfo {author}
  {\bibfnamefont {I.-O.}\ \bibnamefont {Stamatescu}},\ }\href@noop {} {\emph
  {\bibinfo {title} {Decoherence and the appearance of a classical world in
  quantum theory}}}\ (\bibinfo  {publisher} {Springer Science \& Business
  Media},\ \bibinfo {year} {2013})\BibitemShut {NoStop}%
\bibitem [{\citenamefont {Zurek}(2003{\natexlab{a}})}]{quant_class_zurek}%
  \BibitemOpen
  \bibfield  {author} {\bibinfo {author} {\bibfnamefont {W.~H.}\ \bibnamefont
  {Zurek}},\ }\href@noop {} {\bibfield  {journal} {\bibinfo  {journal} {Rev.
  Mod. Phys.}\ }\textbf {\bibinfo {volume} {75}},\ \bibinfo {pages} {715}
  (\bibinfo {year} {2003}{\natexlab{a}})}\BibitemShut {NoStop}%
\bibitem [{\citenamefont {Zavatta}\ \emph {et~al.}(2004)\citenamefont
  {Zavatta}, \citenamefont {Viciani},\ and\ \citenamefont
  {Bellini}}]{quant_class_exp_zavatta}%
  \BibitemOpen
  \bibfield  {author} {\bibinfo {author} {\bibfnamefont {A.}~\bibnamefont
  {Zavatta}}, \bibinfo {author} {\bibfnamefont {S.}~\bibnamefont {Viciani}}, \
  and\ \bibinfo {author} {\bibfnamefont {M.}~\bibnamefont {Bellini}},\
  }\href@noop {} {\bibfield  {journal} {\bibinfo  {journal} {Science}\ }\textbf
  {\bibinfo {volume} {306}},\ \bibinfo {pages} {660} (\bibinfo {year}
  {2004})}\BibitemShut {NoStop}%
\bibitem [{\citenamefont {Modi}\ \emph {et~al.}(2012)\citenamefont {Modi},
  \citenamefont {Brodutch}, \citenamefont {Cable}, \citenamefont {Paterek},\
  and\ \citenamefont {Vedral}}]{quant_class_modi}%
  \BibitemOpen
  \bibfield  {author} {\bibinfo {author} {\bibfnamefont {K.}~\bibnamefont
  {Modi}}, \bibinfo {author} {\bibfnamefont {A.}~\bibnamefont {Brodutch}},
  \bibinfo {author} {\bibfnamefont {H.}~\bibnamefont {Cable}}, \bibinfo
  {author} {\bibfnamefont {T.}~\bibnamefont {Paterek}}, \ and\ \bibinfo
  {author} {\bibfnamefont {V.}~\bibnamefont {Vedral}},\ }\href@noop {}
  {\bibfield  {journal} {\bibinfo  {journal} {Rev. Mod. Phys.}\ }\textbf
  {\bibinfo {volume} {84}},\ \bibinfo {pages} {1655} (\bibinfo {year}
  {2012})}\BibitemShut {NoStop}%
\bibitem [{\citenamefont {Haroche}(2013)}]{quant_class_haroche}%
  \BibitemOpen
  \bibfield  {author} {\bibinfo {author} {\bibfnamefont {S.}~\bibnamefont
  {Haroche}},\ }\href@noop {} {\bibfield  {journal} {\bibinfo  {journal} {Rev.
  Mod. Phys.}\ }\textbf {\bibinfo {volume} {85}},\ \bibinfo {pages} {1083}
  (\bibinfo {year} {2013})}\BibitemShut {NoStop}%
\bibitem [{\citenamefont {Bassi}\ \emph {et~al.}(2013)\citenamefont {Bassi},
  \citenamefont {Lochan}, \citenamefont {Satin}, \citenamefont {Singh},\ and\
  \citenamefont {Ulbricht}}]{collapse_bassi_rmp}%
  \BibitemOpen
  \bibfield  {author} {\bibinfo {author} {\bibfnamefont {A.}~\bibnamefont
  {Bassi}}, \bibinfo {author} {\bibfnamefont {K.}~\bibnamefont {Lochan}},
  \bibinfo {author} {\bibfnamefont {S.}~\bibnamefont {Satin}}, \bibinfo
  {author} {\bibfnamefont {T.~P.}\ \bibnamefont {Singh}}, \ and\ \bibinfo
  {author} {\bibfnamefont {H.}~\bibnamefont {Ulbricht}},\ }\href@noop {}
  {\bibfield  {journal} {\bibinfo  {journal} {Rev. Mod. Phys.}\ }\textbf
  {\bibinfo {volume} {85}},\ \bibinfo {pages} {471} (\bibinfo {year}
  {2013})}\BibitemShut {NoStop}%
\bibitem [{\citenamefont {Romero-Isart}(2011)}]{collapse_romero}%
  \BibitemOpen
  \bibfield  {author} {\bibinfo {author} {\bibfnamefont {O.}~\bibnamefont
  {Romero-Isart}},\ }\href@noop {} {\bibfield  {journal} {\bibinfo  {journal}
  {Physical Review A}\ }\textbf {\bibinfo {volume} {84}},\ \bibinfo {pages}
  {052121} (\bibinfo {year} {2011})}\BibitemShut {NoStop}%
\bibitem [{\citenamefont {Fr{\"o}wis}\ \emph {et~al.}(2018)\citenamefont
  {Fr{\"o}wis}, \citenamefont {Sekatski}, \citenamefont {D{\"u}r},
  \citenamefont {Gisin},\ and\ \citenamefont
  {Sangouard}}]{macroscopic_quantum_review}%
  \BibitemOpen
  \bibfield  {author} {\bibinfo {author} {\bibfnamefont {F.}~\bibnamefont
  {Fr{\"o}wis}}, \bibinfo {author} {\bibfnamefont {P.}~\bibnamefont
  {Sekatski}}, \bibinfo {author} {\bibfnamefont {W.}~\bibnamefont {D{\"u}r}},
  \bibinfo {author} {\bibfnamefont {N.}~\bibnamefont {Gisin}}, \ and\ \bibinfo
  {author} {\bibfnamefont {N.}~\bibnamefont {Sangouard}},\ }\href@noop {}
  {\bibfield  {journal} {\bibinfo  {journal} {Rev. Mod. Phys.}\ }\textbf
  {\bibinfo {volume} {90}},\ \bibinfo {pages} {025004} (\bibinfo {year}
  {2018})}\BibitemShut {NoStop}%
\bibitem [{\citenamefont {Arndt}\ \emph {et~al.}(1999)\citenamefont {Arndt},
  \citenamefont {Nairz}, \citenamefont {Vos-Andreae}, \citenamefont {Keller},
  \citenamefont {Van~der Zouw},\ and\ \citenamefont {Zeilinger}}]{arndt_c60}%
  \BibitemOpen
  \bibfield  {author} {\bibinfo {author} {\bibfnamefont {M.}~\bibnamefont
  {Arndt}}, \bibinfo {author} {\bibfnamefont {O.}~\bibnamefont {Nairz}},
  \bibinfo {author} {\bibfnamefont {J.}~\bibnamefont {Vos-Andreae}}, \bibinfo
  {author} {\bibfnamefont {C.}~\bibnamefont {Keller}}, \bibinfo {author}
  {\bibfnamefont {G.}~\bibnamefont {Van~der Zouw}}, \ and\ \bibinfo {author}
  {\bibfnamefont {A.}~\bibnamefont {Zeilinger}},\ }\href@noop {} {\bibfield
  {journal} {\bibinfo  {journal} {Nature}\ }\textbf {\bibinfo {volume} {401}},\
  \bibinfo {pages} {680} (\bibinfo {year} {1999})}\BibitemShut {NoStop}%
\bibitem [{\citenamefont {Gerlich}\ \emph {et~al.}(2011)\citenamefont
  {Gerlich}, \citenamefont {Eibenberger}, \citenamefont {Tomandl},
  \citenamefont {Nimmrichter}, \citenamefont {Hornberger}, \citenamefont
  {Fagan}, \citenamefont {T{\"u}xen}, \citenamefont {Mayor},\ and\
  \citenamefont {Arndt}}]{organic_gerlich}%
  \BibitemOpen
  \bibfield  {author} {\bibinfo {author} {\bibfnamefont {S.}~\bibnamefont
  {Gerlich}}, \bibinfo {author} {\bibfnamefont {S.}~\bibnamefont
  {Eibenberger}}, \bibinfo {author} {\bibfnamefont {M.}~\bibnamefont
  {Tomandl}}, \bibinfo {author} {\bibfnamefont {S.}~\bibnamefont
  {Nimmrichter}}, \bibinfo {author} {\bibfnamefont {K.}~\bibnamefont
  {Hornberger}}, \bibinfo {author} {\bibfnamefont {P.~J.}\ \bibnamefont
  {Fagan}}, \bibinfo {author} {\bibfnamefont {J.}~\bibnamefont {T{\"u}xen}},
  \bibinfo {author} {\bibfnamefont {M.}~\bibnamefont {Mayor}}, \ and\ \bibinfo
  {author} {\bibfnamefont {M.}~\bibnamefont {Arndt}},\ }\href@noop {}
  {\bibfield  {journal} {\bibinfo  {journal} {Nature communications}\ }\textbf
  {\bibinfo {volume} {2}},\ \bibinfo {pages} {1} (\bibinfo {year}
  {2011})}\BibitemShut {NoStop}%
\bibitem [{\citenamefont {Eibenberger}\ \emph {et~al.}(2013)\citenamefont
  {Eibenberger}, \citenamefont {Gerlich}, \citenamefont {Arndt}, \citenamefont
  {Mayor},\ and\ \citenamefont {T{\"u}xen}}]{mlibrary_arndt}%
  \BibitemOpen
  \bibfield  {author} {\bibinfo {author} {\bibfnamefont {S.}~\bibnamefont
  {Eibenberger}}, \bibinfo {author} {\bibfnamefont {S.}~\bibnamefont
  {Gerlich}}, \bibinfo {author} {\bibfnamefont {M.}~\bibnamefont {Arndt}},
  \bibinfo {author} {\bibfnamefont {M.}~\bibnamefont {Mayor}}, \ and\ \bibinfo
  {author} {\bibfnamefont {J.}~\bibnamefont {T{\"u}xen}},\ }\href@noop {}
  {\bibfield  {journal} {\bibinfo  {journal} {Physical Chemistry Chemical
  Physics}\ }\textbf {\bibinfo {volume} {15}},\ \bibinfo {pages} {14696}
  (\bibinfo {year} {2013})}\BibitemShut {NoStop}%
\bibitem [{\citenamefont {Van Der~Wal}\ \emph {et~al.}(2000)\citenamefont {Van
  Der~Wal}, \citenamefont {Ter~Haar}, \citenamefont {Wilhelm}, \citenamefont
  {Schouten}, \citenamefont {Harmans}, \citenamefont {Orlando}, \citenamefont
  {Lloyd},\ and\ \citenamefont {Mooij}}]{vanderwal_superconducting}%
  \BibitemOpen
  \bibfield  {author} {\bibinfo {author} {\bibfnamefont {C.~H.}\ \bibnamefont
  {Van Der~Wal}}, \bibinfo {author} {\bibfnamefont {A.}~\bibnamefont
  {Ter~Haar}}, \bibinfo {author} {\bibfnamefont {F.}~\bibnamefont {Wilhelm}},
  \bibinfo {author} {\bibfnamefont {R.}~\bibnamefont {Schouten}}, \bibinfo
  {author} {\bibfnamefont {C.}~\bibnamefont {Harmans}}, \bibinfo {author}
  {\bibfnamefont {T.}~\bibnamefont {Orlando}}, \bibinfo {author} {\bibfnamefont
  {S.}~\bibnamefont {Lloyd}}, \ and\ \bibinfo {author} {\bibfnamefont
  {J.}~\bibnamefont {Mooij}},\ }\href@noop {} {\bibfield  {journal} {\bibinfo
  {journal} {Science}\ }\textbf {\bibinfo {volume} {290}},\ \bibinfo {pages}
  {773} (\bibinfo {year} {2000})}\BibitemShut {NoStop}%
\bibitem [{\citenamefont {Kleckner}\ \emph {et~al.}(2008)\citenamefont
  {Kleckner}, \citenamefont {Pikovski}, \citenamefont {Jeffrey}, \citenamefont
  {Ament}, \citenamefont {Eliel}, \citenamefont {Van Den~Brink},\ and\
  \citenamefont {Bouwmeester}}]{opto_dustin}%
  \BibitemOpen
  \bibfield  {author} {\bibinfo {author} {\bibfnamefont {D.}~\bibnamefont
  {Kleckner}}, \bibinfo {author} {\bibfnamefont {I.}~\bibnamefont {Pikovski}},
  \bibinfo {author} {\bibfnamefont {E.}~\bibnamefont {Jeffrey}}, \bibinfo
  {author} {\bibfnamefont {L.}~\bibnamefont {Ament}}, \bibinfo {author}
  {\bibfnamefont {E.}~\bibnamefont {Eliel}}, \bibinfo {author} {\bibfnamefont
  {J.}~\bibnamefont {Van Den~Brink}}, \ and\ \bibinfo {author} {\bibfnamefont
  {D.}~\bibnamefont {Bouwmeester}},\ }\href@noop {} {\bibfield  {journal}
  {\bibinfo  {journal} {New J. Phys.}\ }\textbf {\bibinfo {volume} {10}},\
  \bibinfo {pages} {095020} (\bibinfo {year} {2008})}\BibitemShut {NoStop}%
\bibitem [{\citenamefont {Tara}\ \emph {et~al.}(1993)\citenamefont {Tara},
  \citenamefont {Agarwal},\ and\ \citenamefont {Chaturvedi}}]{macsup_tara}%
  \BibitemOpen
  \bibfield  {author} {\bibinfo {author} {\bibfnamefont {K.}~\bibnamefont
  {Tara}}, \bibinfo {author} {\bibfnamefont {G.}~\bibnamefont {Agarwal}}, \
  and\ \bibinfo {author} {\bibfnamefont {S.}~\bibnamefont {Chaturvedi}},\
  }\href@noop {} {\bibfield  {journal} {\bibinfo  {journal} {Physical Review
  A}\ }\textbf {\bibinfo {volume} {47}},\ \bibinfo {pages} {5024} (\bibinfo
  {year} {1993})}\BibitemShut {NoStop}%
\bibitem [{\citenamefont {M{\"o}bius}\ \emph {et~al.}(2013)\citenamefont
  {M{\"o}bius}, \citenamefont {Genkin}, \citenamefont {Eisfeld}, \citenamefont
  {W{\"u}ster},\ and\ \citenamefont {{J.-M. Rost}}}]{moebius:cat}%
  \BibitemOpen
  \bibfield  {author} {\bibinfo {author} {\bibfnamefont {S.}~\bibnamefont
  {M{\"o}bius}}, \bibinfo {author} {\bibfnamefont {M.}~\bibnamefont {Genkin}},
  \bibinfo {author} {\bibfnamefont {A.}~\bibnamefont {Eisfeld}}, \bibinfo
  {author} {\bibfnamefont {S.}~\bibnamefont {W{\"u}ster}}, \ and\ \bibinfo
  {author} {\bibnamefont {{J.-M. Rost}}},\ }\href@noop {} {\bibfield  {journal}
  {\bibinfo  {journal} {Phys. Rev. A}\ }\textbf {\bibinfo {volume} {87}},\
  \bibinfo {pages} {051602} (\bibinfo {year} {2013})}\BibitemShut {NoStop}%
\bibitem [{\citenamefont {Sreedharan}\ \emph {et~al.}(2020)\citenamefont
  {Sreedharan}, \citenamefont {Choudhury}, \citenamefont {Mukherjee},
  \citenamefont {Streltsov},\ and\ \citenamefont
  {W{\"u}ster}}]{soliton_aparna}%
  \BibitemOpen
  \bibfield  {author} {\bibinfo {author} {\bibfnamefont {A.}~\bibnamefont
  {Sreedharan}}, \bibinfo {author} {\bibfnamefont {S.}~\bibnamefont
  {Choudhury}}, \bibinfo {author} {\bibfnamefont {R.}~\bibnamefont
  {Mukherjee}}, \bibinfo {author} {\bibfnamefont {A.}~\bibnamefont
  {Streltsov}}, \ and\ \bibinfo {author} {\bibfnamefont {S.}~\bibnamefont
  {W{\"u}ster}},\ }\href@noop {} {\bibfield  {journal} {\bibinfo  {journal}
  {Phys. Rev. A}\ }\textbf {\bibinfo {volume} {101}},\ \bibinfo {pages}
  {043604} (\bibinfo {year} {2020})}\BibitemShut {NoStop}%
\bibitem [{\citenamefont {Streltsov}\ \emph {et~al.}(2009)\citenamefont
  {Streltsov}, \citenamefont {Alon},\ and\ \citenamefont
  {Cederbaum}}]{barrier_scattering_superposition}%
  \BibitemOpen
  \bibfield  {author} {\bibinfo {author} {\bibfnamefont {A.~I.}\ \bibnamefont
  {Streltsov}}, \bibinfo {author} {\bibfnamefont {O.~E.}\ \bibnamefont {Alon}},
  \ and\ \bibinfo {author} {\bibfnamefont {L.~S.}\ \bibnamefont {Cederbaum}},\
  }\href@noop {} {\bibfield  {journal} {\bibinfo  {journal} {Physical Review
  A}\ }\textbf {\bibinfo {volume} {80}},\ \bibinfo {pages} {043616} (\bibinfo
  {year} {2009})}\BibitemShut {NoStop}%
\bibitem [{\citenamefont {Mukherjee}\ \emph {et~al.}(2015)\citenamefont
  {Mukherjee}, \citenamefont {Ates}, \citenamefont {Li},\ and\ \citenamefont
  {W{\"u}ster}}]{sebnrick}%
  \BibitemOpen
  \bibfield  {author} {\bibinfo {author} {\bibfnamefont {R.}~\bibnamefont
  {Mukherjee}}, \bibinfo {author} {\bibfnamefont {C.}~\bibnamefont {Ates}},
  \bibinfo {author} {\bibfnamefont {W.}~\bibnamefont {Li}}, \ and\ \bibinfo
  {author} {\bibfnamefont {S.}~\bibnamefont {W{\"u}ster}},\ }\href@noop {}
  {\bibfield  {journal} {\bibinfo  {journal} {Phys. Rev. Lett.}\ }\textbf
  {\bibinfo {volume} {115}},\ \bibinfo {pages} {040401} (\bibinfo {year}
  {2015})}\BibitemShut {NoStop}%
\bibitem [{\citenamefont {Cirac}\ \emph {et~al.}(1998)\citenamefont {Cirac},
  \citenamefont {Lewenstein}, \citenamefont {M{\o}lmer},\ and\ \citenamefont
  {Zoller}}]{bec_superpos_cirac}%
  \BibitemOpen
  \bibfield  {author} {\bibinfo {author} {\bibfnamefont {J.~I.}\ \bibnamefont
  {Cirac}}, \bibinfo {author} {\bibfnamefont {M.}~\bibnamefont {Lewenstein}},
  \bibinfo {author} {\bibfnamefont {K.}~\bibnamefont {M{\o}lmer}}, \ and\
  \bibinfo {author} {\bibfnamefont {P.}~\bibnamefont {Zoller}},\ }\href@noop {}
  {\bibfield  {journal} {\bibinfo  {journal} {Physical Review A}\ }\textbf
  {\bibinfo {volume} {57}},\ \bibinfo {pages} {1208} (\bibinfo {year}
  {1998})}\BibitemShut {NoStop}%
\bibitem [{\citenamefont {Pezze}\ \emph {et~al.}(2019)\citenamefont {Pezze},
  \citenamefont {Gessner}, \citenamefont {Feldmann}, \citenamefont {Klempt},
  \citenamefont {Santos},\ and\ \citenamefont {Smerzi}}]{bec_superpos_pezze}%
  \BibitemOpen
  \bibfield  {author} {\bibinfo {author} {\bibfnamefont {L.}~\bibnamefont
  {Pezze}}, \bibinfo {author} {\bibfnamefont {M.}~\bibnamefont {Gessner}},
  \bibinfo {author} {\bibfnamefont {P.}~\bibnamefont {Feldmann}}, \bibinfo
  {author} {\bibfnamefont {C.}~\bibnamefont {Klempt}}, \bibinfo {author}
  {\bibfnamefont {L.}~\bibnamefont {Santos}}, \ and\ \bibinfo {author}
  {\bibfnamefont {A.}~\bibnamefont {Smerzi}},\ }\href@noop {} {\bibfield
  {journal} {\bibinfo  {journal} {Physical Review Letters}\ }\textbf {\bibinfo
  {volume} {123}},\ \bibinfo {pages} {260403} (\bibinfo {year}
  {2019})}\BibitemShut {NoStop}%
\bibitem [{\citenamefont {Gordon}\ and\ \citenamefont
  {Savage}(1999)}]{gordon:BECcat}%
  \BibitemOpen
  \bibfield  {author} {\bibinfo {author} {\bibfnamefont {D.}~\bibnamefont
  {Gordon}}\ and\ \bibinfo {author} {\bibfnamefont {C.~M.}\ \bibnamefont
  {Savage}},\ }\href@noop {} {\bibfield  {journal} {\bibinfo  {journal} {Phys.
  Rev. A}\ }\textbf {\bibinfo {volume} {59}},\ \bibinfo {pages} {4623}
  (\bibinfo {year} {1999})}\BibitemShut {NoStop}%
\bibitem [{\citenamefont {Weiss}\ and\ \citenamefont
  {Castin}(2009)}]{weiss:solitoncat}%
  \BibitemOpen
  \bibfield  {author} {\bibinfo {author} {\bibfnamefont {C.}~\bibnamefont
  {Weiss}}\ and\ \bibinfo {author} {\bibfnamefont {Y.}~\bibnamefont {Castin}},\
  }\href@noop {} {\bibfield  {journal} {\bibinfo  {journal} {Phys. Rev. Lett.}\
  }\textbf {\bibinfo {volume} {102}},\ \bibinfo {pages} {010403} (\bibinfo
  {year} {2009})}\BibitemShut {NoStop}%
\bibitem [{\citenamefont {Me\ifmmode \check{z}\else \v{z}\fi{}nar\ifmmode
  \check{s}\else \v{s}\fi{}i\ifmmode~\check{c}\else \v{c}\fi{}}\ \emph
  {et~al.}(2019)\citenamefont {Me\ifmmode \check{z}\else \v{z}\fi{}nar\ifmmode
  \check{s}\else \v{s}\fi{}i\ifmmode~\check{c}\else \v{c}\fi{}}, \citenamefont
  {Arh}, \citenamefont {Brence}, \citenamefont {Pi\ifmmode~\check{s}\else
  \v{s}\fi{}ljar}, \citenamefont {Gosar}, \citenamefont {Gosar}, \citenamefont
  {\ifmmode~\check{Z}\else \v{Z}\fi{}itko}, \citenamefont
  {Zupani\ifmmode~\check{c}\else \v{c}\fi{}},\ and\ \citenamefont
  {Jegli\ifmmode~\check{c}\else \v{c}\fi{}}}]{aref1}%
  \BibitemOpen
  \bibfield  {author} {\bibinfo {author} {\bibfnamefont {T.}~\bibnamefont
  {Me\ifmmode \check{z}\else \v{z}\fi{}nar\ifmmode \check{s}\else
  \v{s}\fi{}i\ifmmode~\check{c}\else \v{c}\fi{}}}, \bibinfo {author}
  {\bibfnamefont {T.}~\bibnamefont {Arh}}, \bibinfo {author} {\bibfnamefont
  {J.}~\bibnamefont {Brence}}, \bibinfo {author} {\bibfnamefont
  {J.}~\bibnamefont {Pi\ifmmode~\check{s}\else \v{s}\fi{}ljar}}, \bibinfo
  {author} {\bibfnamefont {K.}~\bibnamefont {Gosar}}, \bibinfo {author}
  {\bibfnamefont {i.~c.~v.}\ \bibnamefont {Gosar}}, \bibinfo {author}
  {\bibfnamefont {R.}~\bibnamefont {\ifmmode~\check{Z}\else \v{Z}\fi{}itko}},
  \bibinfo {author} {\bibfnamefont {E.}~\bibnamefont
  {Zupani\ifmmode~\check{c}\else \v{c}\fi{}}}, \ and\ \bibinfo {author}
  {\bibfnamefont {P.}~\bibnamefont {Jegli\ifmmode~\check{c}\else \v{c}\fi{}}},\
  }\href@noop {} {\bibfield  {journal} {\bibinfo  {journal} {Phys. Rev. A}\
  }\textbf {\bibinfo {volume} {99}},\ \bibinfo {pages} {033625} (\bibinfo
  {year} {2019})}\BibitemShut {NoStop}%
\bibitem [{\citenamefont {Everitt}\ \emph {et~al.}(2017)\citenamefont
  {Everitt}, \citenamefont {Sooriyabandara}, \citenamefont {Guasoni},
  \citenamefont {Wigley}, \citenamefont {Wei}, \citenamefont {McDonald},
  \citenamefont {Hardman}, \citenamefont {Manju}, \citenamefont {Close},
  \citenamefont {Kuhn}, \citenamefont {Szigeti}, \citenamefont {Kivshar},\ and\
  \citenamefont {Robins}}]{aref2}%
  \BibitemOpen
  \bibfield  {author} {\bibinfo {author} {\bibfnamefont {P.~J.}\ \bibnamefont
  {Everitt}}, \bibinfo {author} {\bibfnamefont {M.~A.}\ \bibnamefont
  {Sooriyabandara}}, \bibinfo {author} {\bibfnamefont {M.}~\bibnamefont
  {Guasoni}}, \bibinfo {author} {\bibfnamefont {P.~B.}\ \bibnamefont {Wigley}},
  \bibinfo {author} {\bibfnamefont {C.~H.}\ \bibnamefont {Wei}}, \bibinfo
  {author} {\bibfnamefont {G.~D.}\ \bibnamefont {McDonald}}, \bibinfo {author}
  {\bibfnamefont {K.~S.}\ \bibnamefont {Hardman}}, \bibinfo {author}
  {\bibfnamefont {P.}~\bibnamefont {Manju}}, \bibinfo {author} {\bibfnamefont
  {J.~D.}\ \bibnamefont {Close}}, \bibinfo {author} {\bibfnamefont {C.~C.~N.}\
  \bibnamefont {Kuhn}}, \bibinfo {author} {\bibfnamefont {S.~S.}\ \bibnamefont
  {Szigeti}}, \bibinfo {author} {\bibfnamefont {Y.~S.}\ \bibnamefont
  {Kivshar}}, \ and\ \bibinfo {author} {\bibfnamefont {N.~P.}\ \bibnamefont
  {Robins}},\ }\href {\doibase 10.1103/PhysRevA.96.041601} {\bibfield
  {journal} {\bibinfo  {journal} {Phys. Rev. A}\ }\textbf {\bibinfo {volume}
  {96}},\ \bibinfo {pages} {041601(R)} (\bibinfo {year} {2017})}\BibitemShut
  {NoStop}%
\bibitem [{\citenamefont {McDonald}\ \emph {et~al.}(2014)\citenamefont
  {McDonald}, \citenamefont {Kuhn}, \citenamefont {Hardman}, \citenamefont
  {Bennetts}, \citenamefont {Everitt}, \citenamefont {Altin}, \citenamefont
  {Debs}, \citenamefont {Close},\ and\ \citenamefont {Robins}}]{aref3}%
  \BibitemOpen
  \bibfield  {author} {\bibinfo {author} {\bibfnamefont {G.~D.}\ \bibnamefont
  {McDonald}}, \bibinfo {author} {\bibfnamefont {C.~C.~N.}\ \bibnamefont
  {Kuhn}}, \bibinfo {author} {\bibfnamefont {K.~S.}\ \bibnamefont {Hardman}},
  \bibinfo {author} {\bibfnamefont {S.}~\bibnamefont {Bennetts}}, \bibinfo
  {author} {\bibfnamefont {P.~J.}\ \bibnamefont {Everitt}}, \bibinfo {author}
  {\bibfnamefont {P.~A.}\ \bibnamefont {Altin}}, \bibinfo {author}
  {\bibfnamefont {J.~E.}\ \bibnamefont {Debs}}, \bibinfo {author}
  {\bibfnamefont {J.~D.}\ \bibnamefont {Close}}, \ and\ \bibinfo {author}
  {\bibfnamefont {N.~P.}\ \bibnamefont {Robins}},\ }\href {\doibase
  10.1103/PhysRevLett.113.013002} {\bibfield  {journal} {\bibinfo  {journal}
  {Phys. Rev. Lett.}\ }\textbf {\bibinfo {volume} {113}},\ \bibinfo {pages}
  {013002} (\bibinfo {year} {2014})}\BibitemShut {NoStop}%
\bibitem [{\citenamefont {Lepoutre}\ \emph {et~al.}(2016)\citenamefont
  {Lepoutre}, \citenamefont {Fouch\'e}, \citenamefont {Boiss\'e}, \citenamefont
  {Berthet}, \citenamefont {Salomon}, \citenamefont {Aspect},\ and\
  \citenamefont {Bourdel}}]{aref4}%
  \BibitemOpen
  \bibfield  {author} {\bibinfo {author} {\bibfnamefont {S.}~\bibnamefont
  {Lepoutre}}, \bibinfo {author} {\bibfnamefont {L.}~\bibnamefont {Fouch\'e}},
  \bibinfo {author} {\bibfnamefont {A.}~\bibnamefont {Boiss\'e}}, \bibinfo
  {author} {\bibfnamefont {G.}~\bibnamefont {Berthet}}, \bibinfo {author}
  {\bibfnamefont {G.}~\bibnamefont {Salomon}}, \bibinfo {author} {\bibfnamefont
  {A.}~\bibnamefont {Aspect}}, \ and\ \bibinfo {author} {\bibfnamefont
  {T.}~\bibnamefont {Bourdel}},\ }\href@noop {} {\bibfield  {journal} {\bibinfo
   {journal} {Phys. Rev. A}\ }\textbf {\bibinfo {volume} {94}},\ \bibinfo
  {pages} {053626} (\bibinfo {year} {2016})}\BibitemShut {NoStop}%
\bibitem [{\citenamefont {Medley}\ \emph {et~al.}(2014)\citenamefont {Medley},
  \citenamefont {Minar}, \citenamefont {Cizek}, \citenamefont {Berryrieser},\
  and\ \citenamefont {Kasevich}}]{aref5}%
  \BibitemOpen
  \bibfield  {author} {\bibinfo {author} {\bibfnamefont {P.}~\bibnamefont
  {Medley}}, \bibinfo {author} {\bibfnamefont {M.~A.}\ \bibnamefont {Minar}},
  \bibinfo {author} {\bibfnamefont {N.~C.}\ \bibnamefont {Cizek}}, \bibinfo
  {author} {\bibfnamefont {D.}~\bibnamefont {Berryrieser}}, \ and\ \bibinfo
  {author} {\bibfnamefont {M.~A.}\ \bibnamefont {Kasevich}},\ }\href@noop {}
  {\bibfield  {journal} {\bibinfo  {journal} {Phys. Rev. Lett.}\ }\textbf
  {\bibinfo {volume} {112}},\ \bibinfo {pages} {060401} (\bibinfo {year}
  {2014})}\BibitemShut {NoStop}%
\bibitem [{\citenamefont {Marchant}\ \emph {et~al.}(2013)\citenamefont
  {Marchant}, \citenamefont {Billam}, \citenamefont {Wiles}, \citenamefont
  {Yu}, \citenamefont {Gardiner},\ and\ \citenamefont {Cornish}}]{aref6}%
  \BibitemOpen
  \bibfield  {author} {\bibinfo {author} {\bibfnamefont {A.~L.}\ \bibnamefont
  {Marchant}}, \bibinfo {author} {\bibfnamefont {T.~P.}\ \bibnamefont
  {Billam}}, \bibinfo {author} {\bibfnamefont {T.~P.}\ \bibnamefont {Wiles}},
  \bibinfo {author} {\bibfnamefont {M.~M.~H.}\ \bibnamefont {Yu}}, \bibinfo
  {author} {\bibfnamefont {S.~A.}\ \bibnamefont {Gardiner}}, \ and\ \bibinfo
  {author} {\bibfnamefont {S.~L.}\ \bibnamefont {Cornish}},\ }\href@noop {}
  {\bibfield  {journal} {\bibinfo  {journal} {Nature Comm.}\ }\textbf {\bibinfo
  {volume} {4}},\ \bibinfo {pages} {1865} (\bibinfo {year} {2013})}\BibitemShut
  {NoStop}%
\bibitem [{\citenamefont {Marchant}\ \emph {et~al.}(2016)\citenamefont
  {Marchant}, \citenamefont {Billam}, \citenamefont {Yu}, \citenamefont
  {Rakonjac}, \citenamefont {Helm}, \citenamefont {Polo}, \citenamefont
  {Weiss}, \citenamefont {Gardiner},\ and\ \citenamefont {Cornish}}]{aref7}%
  \BibitemOpen
  \bibfield  {author} {\bibinfo {author} {\bibfnamefont {A.~L.}\ \bibnamefont
  {Marchant}}, \bibinfo {author} {\bibfnamefont {T.~P.}\ \bibnamefont
  {Billam}}, \bibinfo {author} {\bibfnamefont {M.~M.~H.}\ \bibnamefont {Yu}},
  \bibinfo {author} {\bibfnamefont {A.}~\bibnamefont {Rakonjac}}, \bibinfo
  {author} {\bibfnamefont {J.~L.}\ \bibnamefont {Helm}}, \bibinfo {author}
  {\bibfnamefont {J.}~\bibnamefont {Polo}}, \bibinfo {author} {\bibfnamefont
  {C.}~\bibnamefont {Weiss}}, \bibinfo {author} {\bibfnamefont {S.~A.}\
  \bibnamefont {Gardiner}}, \ and\ \bibinfo {author} {\bibfnamefont {S.~L.}\
  \bibnamefont {Cornish}},\ }\href {\doibase 10.1103/PhysRevA.93.021604}
  {\bibfield  {journal} {\bibinfo  {journal} {Phys. Rev. A}\ }\textbf {\bibinfo
  {volume} {93}},\ \bibinfo {pages} {021604(R)} (\bibinfo {year}
  {2016})}\BibitemShut {NoStop}%
\bibitem [{\citenamefont {Nguyen}\ \emph {et~al.}(2014)\citenamefont {Nguyen},
  \citenamefont {Dyke}, \citenamefont {Luo}, \citenamefont {Malomed},\ and\
  \citenamefont {Hulet}}]{aref8}%
  \BibitemOpen
  \bibfield  {author} {\bibinfo {author} {\bibfnamefont {J.~H.~V.}\
  \bibnamefont {Nguyen}}, \bibinfo {author} {\bibfnamefont {P.}~\bibnamefont
  {Dyke}}, \bibinfo {author} {\bibfnamefont {D.}~\bibnamefont {Luo}}, \bibinfo
  {author} {\bibfnamefont {B.~A.}\ \bibnamefont {Malomed}}, \ and\ \bibinfo
  {author} {\bibfnamefont {R.~G.}\ \bibnamefont {Hulet}},\ }\href@noop {}
  {\bibfield  {journal} {\bibinfo  {journal} {Nature Physics}\ }\textbf
  {\bibinfo {volume} {10}},\ \bibinfo {pages} {918} (\bibinfo {year}
  {2014})}\BibitemShut {NoStop}%
\bibitem [{\citenamefont {Nguyen}\ \emph {et~al.}(2017)\citenamefont {Nguyen},
  \citenamefont {Luo},\ and\ \citenamefont {Hulet}}]{aref9}%
  \BibitemOpen
  \bibfield  {author} {\bibinfo {author} {\bibfnamefont {J.~H.~V.}\
  \bibnamefont {Nguyen}}, \bibinfo {author} {\bibfnamefont {D.}~\bibnamefont
  {Luo}}, \ and\ \bibinfo {author} {\bibfnamefont {R.~G.}\ \bibnamefont
  {Hulet}},\ }\href@noop {} {\bibfield  {journal} {\bibinfo  {journal}
  {Science}\ }\textbf {\bibinfo {volume} {356}},\ \bibinfo {pages} {422}
  (\bibinfo {year} {2017})}\BibitemShut {NoStop}%
\bibitem [{\citenamefont {Cornish}\ \emph {et~al.}(2006)\citenamefont
  {Cornish}, \citenamefont {Thompson},\ and\ \citenamefont {Wieman}}]{aref10}%
  \BibitemOpen
  \bibfield  {author} {\bibinfo {author} {\bibfnamefont {S.~L.}\ \bibnamefont
  {Cornish}}, \bibinfo {author} {\bibfnamefont {S.~T.}\ \bibnamefont
  {Thompson}}, \ and\ \bibinfo {author} {\bibfnamefont {C.~E.}\ \bibnamefont
  {Wieman}},\ }\href@noop {} {\bibfield  {journal} {\bibinfo  {journal} {Phys.
  Rev. Lett.}\ }\textbf {\bibinfo {volume} {96}},\ \bibinfo {pages} {170401}
  (\bibinfo {year} {2006})}\BibitemShut {NoStop}%
\bibitem [{\citenamefont {Eiermann}\ \emph {et~al.}(2004)\citenamefont
  {Eiermann}, \citenamefont {Anker}, \citenamefont {Albiez}, \citenamefont
  {Taglieber}, \citenamefont {Treutlein}, \citenamefont {Marzlin},\ and\
  \citenamefont {Oberthaler}}]{aref11}%
  \BibitemOpen
  \bibfield  {author} {\bibinfo {author} {\bibfnamefont {B.}~\bibnamefont
  {Eiermann}}, \bibinfo {author} {\bibfnamefont {T.}~\bibnamefont {Anker}},
  \bibinfo {author} {\bibfnamefont {M.}~\bibnamefont {Albiez}}, \bibinfo
  {author} {\bibfnamefont {M.}~\bibnamefont {Taglieber}}, \bibinfo {author}
  {\bibfnamefont {P.}~\bibnamefont {Treutlein}}, \bibinfo {author}
  {\bibfnamefont {K.-P.}\ \bibnamefont {Marzlin}}, \ and\ \bibinfo {author}
  {\bibfnamefont {M.~K.}\ \bibnamefont {Oberthaler}},\ }\href {\doibase
  10.1103/PhysRevLett.92.230401} {\bibfield  {journal} {\bibinfo  {journal}
  {Phys. Rev. Lett.}\ }\textbf {\bibinfo {volume} {92}},\ \bibinfo {pages}
  {230401} (\bibinfo {year} {2004})}\BibitemShut {NoStop}%
\bibitem [{\citenamefont {Strecker}\ \emph {et~al.}(2002)\citenamefont
  {Strecker}, \citenamefont {Partridge}, \citenamefont {Truscott},\ and\
  \citenamefont {Hulet}}]{aref12}%
  \BibitemOpen
  \bibfield  {author} {\bibinfo {author} {\bibfnamefont {K.~E.}\ \bibnamefont
  {Strecker}}, \bibinfo {author} {\bibfnamefont {G.~B.}\ \bibnamefont
  {Partridge}}, \bibinfo {author} {\bibfnamefont {A.~G.}\ \bibnamefont
  {Truscott}}, \ and\ \bibinfo {author} {\bibfnamefont {R.~G.}\ \bibnamefont
  {Hulet}},\ }\href@noop {} {\bibfield  {journal} {\bibinfo  {journal}
  {Nature}\ }\textbf {\bibinfo {volume} {417}},\ \bibinfo {pages} {150}
  (\bibinfo {year} {2002})}\BibitemShut {NoStop}%
\bibitem [{\citenamefont {Khaykovich}\ \emph {et~al.}(2002)\citenamefont
  {Khaykovich}, \citenamefont {Schreck}, \citenamefont {Ferrari}, \citenamefont
  {Bourdel}, \citenamefont {Cubizolles}, \citenamefont {Carr}, \citenamefont
  {Castin},\ and\ \citenamefont {Salomon}}]{aref13}%
  \BibitemOpen
  \bibfield  {author} {\bibinfo {author} {\bibfnamefont {L.}~\bibnamefont
  {Khaykovich}}, \bibinfo {author} {\bibfnamefont {F.}~\bibnamefont {Schreck}},
  \bibinfo {author} {\bibfnamefont {G.}~\bibnamefont {Ferrari}}, \bibinfo
  {author} {\bibfnamefont {T.}~\bibnamefont {Bourdel}}, \bibinfo {author}
  {\bibfnamefont {J.}~\bibnamefont {Cubizolles}}, \bibinfo {author}
  {\bibfnamefont {L.~D.}\ \bibnamefont {Carr}}, \bibinfo {author}
  {\bibfnamefont {Y.}~\bibnamefont {Castin}}, \ and\ \bibinfo {author}
  {\bibfnamefont {C.}~\bibnamefont {Salomon}},\ }\href {\doibase
  10.1126/science.1071021} {\bibfield  {journal} {\bibinfo  {journal}
  {Science}\ }\textbf {\bibinfo {volume} {296}},\ \bibinfo {pages} {1290}
  (\bibinfo {year} {2002})}\BibitemShut {NoStop}%
\bibitem [{\citenamefont {Kivshar}\ and\ \citenamefont
  {Agrawal}(2003)}]{aref14}%
  \BibitemOpen
  \bibfield  {author} {\bibinfo {author} {\bibfnamefont {Y.~S.}\ \bibnamefont
  {Kivshar}}\ and\ \bibinfo {author} {\bibfnamefont {G.~P.}\ \bibnamefont
  {Agrawal}},\ }\href@noop {} {\emph {\bibinfo {title} {Optical Solitons: From
  Fibers to Photonic Crystals}}}\ (\bibinfo  {publisher} {Academic, San
  Diego},\ \bibinfo {year} {2003})\BibitemShut {NoStop}%
\bibitem [{\citenamefont {Pethick}\ and\ \citenamefont {Smith}(2002)}]{aref15}%
  \BibitemOpen
  \bibfield  {author} {\bibinfo {author} {\bibfnamefont {C.~J.}\ \bibnamefont
  {Pethick}}\ and\ \bibinfo {author} {\bibfnamefont {H.}~\bibnamefont
  {Smith}},\ }\href@noop {} {\emph {\bibinfo {title} {{Bose-Einstein}
  condensation in dilute gases}}}\ (\bibinfo  {publisher} {Cambridge University
  Press},\ \bibinfo {year} {2002})\BibitemShut {NoStop}%
\bibitem [{\citenamefont {Strecker}\ \emph {et~al.}(2003)\citenamefont
  {Strecker}, \citenamefont {Partridge}, \citenamefont {Truscott},\ and\
  \citenamefont {Hulet}}]{aref16}%
  \BibitemOpen
  \bibfield  {author} {\bibinfo {author} {\bibfnamefont {K.~E.}\ \bibnamefont
  {Strecker}}, \bibinfo {author} {\bibfnamefont {G.~B.}\ \bibnamefont
  {Partridge}}, \bibinfo {author} {\bibfnamefont {A.~G.}\ \bibnamefont
  {Truscott}}, \ and\ \bibinfo {author} {\bibfnamefont {R.~G.}\ \bibnamefont
  {Hulet}},\ }\href@noop {} {\bibfield  {journal} {\bibinfo  {journal} {New
  Journal of Physics}\ }\textbf {\bibinfo {volume} {5}},\ \bibinfo {pages} {73}
  (\bibinfo {year} {2003})}\BibitemShut {NoStop}%
\bibitem [{\citenamefont {Gallagher}(2005)}]{gallagher_Rydberg_book}%
  \BibitemOpen
  \bibfield  {author} {\bibinfo {author} {\bibfnamefont {T.~F.}\ \bibnamefont
  {Gallagher}},\ }\href@noop {} {\emph {\bibinfo {title} {Rydberg atoms}}},\
  Vol.~\bibinfo {volume} {3}\ (\bibinfo  {publisher} {Cambridge University
  Press},\ \bibinfo {year} {2005})\BibitemShut {NoStop}%
\bibitem [{\citenamefont {Gallagher}(1988)}]{gallagher_Rydberg_review}%
  \BibitemOpen
  \bibfield  {author} {\bibinfo {author} {\bibfnamefont {T.}~\bibnamefont
  {Gallagher}},\ }\href@noop {} {\bibfield  {journal} {\bibinfo  {journal}
  {Reports on Progress in Physics}\ }\textbf {\bibinfo {volume} {51}},\
  \bibinfo {pages} {143} (\bibinfo {year} {1988})}\BibitemShut {NoStop}%
\bibitem [{\citenamefont {Ga{\"e}tan}\ \emph {et~al.}(2009)\citenamefont
  {Ga{\"e}tan}, \citenamefont {Miroshnychenko}, \citenamefont {Wilk},
  \citenamefont {Chotia}, \citenamefont {Viteau}, \citenamefont {Comparat},
  \citenamefont {Pillet}, \citenamefont {Browaeys},\ and\ \citenamefont
  {Grangier}}]{gaetan:twoatomblock}%
  \BibitemOpen
  \bibfield  {author} {\bibinfo {author} {\bibfnamefont {A.}~\bibnamefont
  {Ga{\"e}tan}}, \bibinfo {author} {\bibfnamefont {Y.}~\bibnamefont
  {Miroshnychenko}}, \bibinfo {author} {\bibfnamefont {T.}~\bibnamefont
  {Wilk}}, \bibinfo {author} {\bibfnamefont {A.}~\bibnamefont {Chotia}},
  \bibinfo {author} {\bibfnamefont {M.}~\bibnamefont {Viteau}}, \bibinfo
  {author} {\bibfnamefont {D.}~\bibnamefont {Comparat}}, \bibinfo {author}
  {\bibfnamefont {P.}~\bibnamefont {Pillet}}, \bibinfo {author} {\bibfnamefont
  {A.}~\bibnamefont {Browaeys}}, \ and\ \bibinfo {author} {\bibfnamefont
  {P.}~\bibnamefont {Grangier}},\ }\href {\doibase 10.1038/nphys1183}
  {\bibfield  {journal} {\bibinfo  {journal} {Nature Physics}\ }\textbf
  {\bibinfo {volume} {5}},\ \bibinfo {pages} {115} (\bibinfo {year}
  {2009})}\BibitemShut {NoStop}%
\bibitem [{\citenamefont {Urban}\ \emph {et~al.}(2009)\citenamefont {Urban},
  \citenamefont {Johnson}, \citenamefont {Henage}, \citenamefont {Isenhower},
  \citenamefont {Yavuz}, \citenamefont {Walker},\ and\ \citenamefont
  {Saffman}}]{rblockade_exp_urban}%
  \BibitemOpen
  \bibfield  {author} {\bibinfo {author} {\bibfnamefont {E.}~\bibnamefont
  {Urban}}, \bibinfo {author} {\bibfnamefont {T.~A.}\ \bibnamefont {Johnson}},
  \bibinfo {author} {\bibfnamefont {T.}~\bibnamefont {Henage}}, \bibinfo
  {author} {\bibfnamefont {L.}~\bibnamefont {Isenhower}}, \bibinfo {author}
  {\bibfnamefont {D.}~\bibnamefont {Yavuz}}, \bibinfo {author} {\bibfnamefont
  {T.}~\bibnamefont {Walker}}, \ and\ \bibinfo {author} {\bibfnamefont
  {M.}~\bibnamefont {Saffman}},\ }\href@noop {} {\bibfield  {journal} {\bibinfo
   {journal} {Nature Physics}\ }\textbf {\bibinfo {volume} {5}},\ \bibinfo
  {pages} {110} (\bibinfo {year} {2009})}\BibitemShut {NoStop}%
\bibitem [{\citenamefont {Jau}\ \emph {et~al.}(2016)\citenamefont {Jau},
  \citenamefont {Hankin}, \citenamefont {Keating}, \citenamefont {Deutsch},\
  and\ \citenamefont {Biedermann}}]{dressing_jau}%
  \BibitemOpen
  \bibfield  {author} {\bibinfo {author} {\bibfnamefont {Y.-Y.}\ \bibnamefont
  {Jau}}, \bibinfo {author} {\bibfnamefont {A.}~\bibnamefont {Hankin}},
  \bibinfo {author} {\bibfnamefont {T.}~\bibnamefont {Keating}}, \bibinfo
  {author} {\bibfnamefont {I.}~\bibnamefont {Deutsch}}, \ and\ \bibinfo
  {author} {\bibfnamefont {G.}~\bibnamefont {Biedermann}},\ }\href@noop {}
  {\bibfield  {journal} {\bibinfo  {journal} {Nature Physics}\ }\textbf
  {\bibinfo {volume} {12}},\ \bibinfo {pages} {71} (\bibinfo {year}
  {2016})}\BibitemShut {NoStop}%
\bibitem [{\citenamefont {Balewski}\ \emph {et~al.}(2014)\citenamefont
  {Balewski}, \citenamefont {Krupp}, \citenamefont {Gaj}, \citenamefont
  {Hofferberth}, \citenamefont {L{\"o}w},\ and\ \citenamefont
  {Pfau}}]{dressing_balewski}%
  \BibitemOpen
  \bibfield  {author} {\bibinfo {author} {\bibfnamefont {J.~B.}\ \bibnamefont
  {Balewski}}, \bibinfo {author} {\bibfnamefont {A.~T.}\ \bibnamefont {Krupp}},
  \bibinfo {author} {\bibfnamefont {A.}~\bibnamefont {Gaj}}, \bibinfo {author}
  {\bibfnamefont {S.}~\bibnamefont {Hofferberth}}, \bibinfo {author}
  {\bibfnamefont {R.}~\bibnamefont {L{\"o}w}}, \ and\ \bibinfo {author}
  {\bibfnamefont {T.}~\bibnamefont {Pfau}},\ }\href@noop {} {\bibfield
  {journal} {\bibinfo  {journal} {New J. Phys.}\ }\textbf {\bibinfo {volume}
  {16}},\ \bibinfo {pages} {063012} (\bibinfo {year} {2014})}\BibitemShut
  {NoStop}%
\bibitem [{\citenamefont {Johnson}\ and\ \citenamefont
  {Rolston}(2010)}]{dressing_johnson}%
  \BibitemOpen
  \bibfield  {author} {\bibinfo {author} {\bibfnamefont {J.}~\bibnamefont
  {Johnson}}\ and\ \bibinfo {author} {\bibfnamefont {S.}~\bibnamefont
  {Rolston}},\ }\href@noop {} {\bibfield  {journal} {\bibinfo  {journal}
  {Physical Review A}\ }\textbf {\bibinfo {volume} {82}},\ \bibinfo {pages}
  {033412} (\bibinfo {year} {2010})}\BibitemShut {NoStop}%
\bibitem [{\citenamefont {W{\"u}ster}\ \emph {et~al.}(2011)\citenamefont
  {W{\"u}ster}, \citenamefont {Ates}, \citenamefont {Eisfeld},\ and\
  \citenamefont {Rost}}]{dressing_wuester}%
  \BibitemOpen
  \bibfield  {author} {\bibinfo {author} {\bibfnamefont {S.}~\bibnamefont
  {W{\"u}ster}}, \bibinfo {author} {\bibfnamefont {C.}~\bibnamefont {Ates}},
  \bibinfo {author} {\bibfnamefont {A.}~\bibnamefont {Eisfeld}}, \ and\
  \bibinfo {author} {\bibfnamefont {J.}~\bibnamefont {Rost}},\ }\href@noop {}
  {\bibfield  {journal} {\bibinfo  {journal} {New J. Phys.}\ }\textbf {\bibinfo
  {volume} {13}},\ \bibinfo {pages} {073044} (\bibinfo {year}
  {2011})}\BibitemShut {NoStop}%
\bibitem [{\citenamefont {Henkel}\ \emph {et~al.}(2010)\citenamefont {Henkel},
  \citenamefont {Nath},\ and\ \citenamefont {Pohl}}]{dressing_henkel}%
  \BibitemOpen
  \bibfield  {author} {\bibinfo {author} {\bibfnamefont {N.}~\bibnamefont
  {Henkel}}, \bibinfo {author} {\bibfnamefont {R.}~\bibnamefont {Nath}}, \ and\
  \bibinfo {author} {\bibfnamefont {T.}~\bibnamefont {Pohl}},\ }\href@noop {}
  {\bibfield  {journal} {\bibinfo  {journal} {Phys. Rev. Lett.}\ }\textbf
  {\bibinfo {volume} {104}},\ \bibinfo {pages} {195302} (\bibinfo {year}
  {2010})}\BibitemShut {NoStop}%
\bibitem [{\citenamefont {Santos}\ \emph {et~al.}(2000)\citenamefont {Santos},
  \citenamefont {Shlyapnikov}, \citenamefont {Zoller},\ and\ \citenamefont
  {Lewenstein}}]{dressing_santos}%
  \BibitemOpen
  \bibfield  {author} {\bibinfo {author} {\bibfnamefont {L.}~\bibnamefont
  {Santos}}, \bibinfo {author} {\bibfnamefont {G.}~\bibnamefont {Shlyapnikov}},
  \bibinfo {author} {\bibfnamefont {P.}~\bibnamefont {Zoller}}, \ and\ \bibinfo
  {author} {\bibfnamefont {M.}~\bibnamefont {Lewenstein}},\ }\href@noop {}
  {\bibfield  {journal} {\bibinfo  {journal} {Physical Review Letters}\
  }\textbf {\bibinfo {volume} {85}},\ \bibinfo {pages} {1791} (\bibinfo {year}
  {2000})}\BibitemShut {NoStop}%
\bibitem [{\citenamefont {Maucher}\ \emph {et~al.}(2011)\citenamefont
  {Maucher}, \citenamefont {Henkel}, \citenamefont {Saffman}, \citenamefont
  {Kr{\'o}likowski}, \citenamefont {Skupin},\ and\ \citenamefont
  {Pohl}}]{dressing_maucher}%
  \BibitemOpen
  \bibfield  {author} {\bibinfo {author} {\bibfnamefont {F.}~\bibnamefont
  {Maucher}}, \bibinfo {author} {\bibfnamefont {N.}~\bibnamefont {Henkel}},
  \bibinfo {author} {\bibfnamefont {M.}~\bibnamefont {Saffman}}, \bibinfo
  {author} {\bibfnamefont {W.}~\bibnamefont {Kr{\'o}likowski}}, \bibinfo
  {author} {\bibfnamefont {S.}~\bibnamefont {Skupin}}, \ and\ \bibinfo {author}
  {\bibfnamefont {T.}~\bibnamefont {Pohl}},\ }\href@noop {} {\bibfield
  {journal} {\bibinfo  {journal} {Phys. Rev. Lett.}\ }\textbf {\bibinfo
  {volume} {106}},\ \bibinfo {pages} {170401} (\bibinfo {year}
  {2011})}\BibitemShut {NoStop}%
\bibitem [{\citenamefont {Honer}\ \emph {et~al.}(2010)\citenamefont {Honer},
  \citenamefont {Weimer}, \citenamefont {Pfau},\ and\ \citenamefont
  {B{\"u}chler}}]{dressing_honer}%
  \BibitemOpen
  \bibfield  {author} {\bibinfo {author} {\bibfnamefont {J.}~\bibnamefont
  {Honer}}, \bibinfo {author} {\bibfnamefont {H.}~\bibnamefont {Weimer}},
  \bibinfo {author} {\bibfnamefont {T.}~\bibnamefont {Pfau}}, \ and\ \bibinfo
  {author} {\bibfnamefont {H.~P.}\ \bibnamefont {B{\"u}chler}},\ }\href@noop {}
  {\bibfield  {journal} {\bibinfo  {journal} {Phys. Rev. Lett.}\ }\textbf
  {\bibinfo {volume} {105}},\ \bibinfo {pages} {160404} (\bibinfo {year}
  {2010})}\BibitemShut {NoStop}%
\bibitem [{\citenamefont {Pupillo}\ \emph {et~al.}(2010)\citenamefont
  {Pupillo}, \citenamefont {Micheli}, \citenamefont {Boninsegni}, \citenamefont
  {Lesanovsky},\ and\ \citenamefont {Zoller}}]{dressing_pupillo}%
  \BibitemOpen
  \bibfield  {author} {\bibinfo {author} {\bibfnamefont {G.}~\bibnamefont
  {Pupillo}}, \bibinfo {author} {\bibfnamefont {A.}~\bibnamefont {Micheli}},
  \bibinfo {author} {\bibfnamefont {M.}~\bibnamefont {Boninsegni}}, \bibinfo
  {author} {\bibfnamefont {I.}~\bibnamefont {Lesanovsky}}, \ and\ \bibinfo
  {author} {\bibfnamefont {P.}~\bibnamefont {Zoller}},\ }\href@noop {}
  {\bibfield  {journal} {\bibinfo  {journal} {Phys. Rev. Lett.}\ }\textbf
  {\bibinfo {volume} {104}},\ \bibinfo {pages} {223002} (\bibinfo {year}
  {2010})}\BibitemShut {NoStop}%
\bibitem [{\citenamefont {Kr{\"o}nke}\ \emph {et~al.}(2013)\citenamefont
  {Kr{\"o}nke}, \citenamefont {Cao}, \citenamefont {Vendrell},\ and\
  \citenamefont {Schmelcher}}]{mlmctdhb_kronke2013}%
  \BibitemOpen
  \bibfield  {author} {\bibinfo {author} {\bibfnamefont {S.}~\bibnamefont
  {Kr{\"o}nke}}, \bibinfo {author} {\bibfnamefont {L.}~\bibnamefont {Cao}},
  \bibinfo {author} {\bibfnamefont {O.}~\bibnamefont {Vendrell}}, \ and\
  \bibinfo {author} {\bibfnamefont {P.}~\bibnamefont {Schmelcher}},\
  }\href@noop {} {\bibfield  {journal} {\bibinfo  {journal} {New J. Phys.}\
  }\textbf {\bibinfo {volume} {15}},\ \bibinfo {pages} {063018} (\bibinfo
  {year} {2013})}\BibitemShut {NoStop}%
\bibitem [{\citenamefont {Schmitz}\ \emph {et~al.}(2013)\citenamefont
  {Schmitz}, \citenamefont {Kr{\"o}nke}, \citenamefont {Cao},\ and\
  \citenamefont {Schmelcher}}]{mlmctdhb_pra}%
  \BibitemOpen
  \bibfield  {author} {\bibinfo {author} {\bibfnamefont {R.}~\bibnamefont
  {Schmitz}}, \bibinfo {author} {\bibfnamefont {S.}~\bibnamefont {Kr{\"o}nke}},
  \bibinfo {author} {\bibfnamefont {L.}~\bibnamefont {Cao}}, \ and\ \bibinfo
  {author} {\bibfnamefont {P.}~\bibnamefont {Schmelcher}},\ }\href@noop {}
  {\bibfield  {journal} {\bibinfo  {journal} {Physical Review A}\ }\textbf
  {\bibinfo {volume} {88}},\ \bibinfo {pages} {043601} (\bibinfo {year}
  {2013})}\BibitemShut {NoStop}%
\bibitem [{\citenamefont {Cao}\ \emph {et~al.}(2013)\citenamefont {Cao},
  \citenamefont {Kr{\"o}nke}, \citenamefont {Vendrell},\ and\ \citenamefont
  {Schmelcher}}]{mlmctdhb_jcp}%
  \BibitemOpen
  \bibfield  {author} {\bibinfo {author} {\bibfnamefont {L.}~\bibnamefont
  {Cao}}, \bibinfo {author} {\bibfnamefont {S.}~\bibnamefont {Kr{\"o}nke}},
  \bibinfo {author} {\bibfnamefont {O.}~\bibnamefont {Vendrell}}, \ and\
  \bibinfo {author} {\bibfnamefont {P.}~\bibnamefont {Schmelcher}},\
  }\href@noop {} {\bibfield  {journal} {\bibinfo  {journal} {J. Chem. Phys.}\
  }\textbf {\bibinfo {volume} {139}},\ \bibinfo {pages} {134103} (\bibinfo
  {year} {2013})}\BibitemShut {NoStop}%
\bibitem [{\citenamefont {Schurer}\ \emph {et~al.}(2014)\citenamefont
  {Schurer}, \citenamefont {Schmelcher},\ and\ \citenamefont
  {Negretti}}]{mlmctdhb_ion_impurity}%
  \BibitemOpen
  \bibfield  {author} {\bibinfo {author} {\bibfnamefont {J.}~\bibnamefont
  {Schurer}}, \bibinfo {author} {\bibfnamefont {P.}~\bibnamefont {Schmelcher}},
  \ and\ \bibinfo {author} {\bibfnamefont {A.}~\bibnamefont {Negretti}},\
  }\href@noop {} {\bibfield  {journal} {\bibinfo  {journal} {Physical Review
  A}\ }\textbf {\bibinfo {volume} {90}},\ \bibinfo {pages} {033601} (\bibinfo
  {year} {2014})}\BibitemShut {NoStop}%
\bibitem [{\citenamefont {Ebgha}\ \emph {et~al.}(2019)\citenamefont {Ebgha},
  \citenamefont {Saeidian}, \citenamefont {Schmelcher},\ and\ \citenamefont
  {Negretti}}]{Ebgha_compoundatomion_PhysRevA}%
  \BibitemOpen
  \bibfield  {author} {\bibinfo {author} {\bibfnamefont {M.~R.}\ \bibnamefont
  {Ebgha}}, \bibinfo {author} {\bibfnamefont {S.}~\bibnamefont {Saeidian}},
  \bibinfo {author} {\bibfnamefont {P.}~\bibnamefont {Schmelcher}}, \ and\
  \bibinfo {author} {\bibfnamefont {A.}~\bibnamefont {Negretti}},\ }\href@noop
  {} {\bibfield  {journal} {\bibinfo  {journal} {Phys. Rev. A}\ }\textbf
  {\bibinfo {volume} {100}},\ \bibinfo {pages} {033616} (\bibinfo {year}
  {2019})}\BibitemShut {NoStop}%
\bibitem [{\citenamefont {Burt}\ \emph {et~al.}(1997)\citenamefont {Burt},
  \citenamefont {Ghrist}, \citenamefont {Myatt}, \citenamefont {Holland},
  \citenamefont {Cornell},\ and\ \citenamefont {Wieman}}]{k1_burt}%
  \BibitemOpen
  \bibfield  {author} {\bibinfo {author} {\bibfnamefont {E.}~\bibnamefont
  {Burt}}, \bibinfo {author} {\bibfnamefont {R.}~\bibnamefont {Ghrist}},
  \bibinfo {author} {\bibfnamefont {C.}~\bibnamefont {Myatt}}, \bibinfo
  {author} {\bibfnamefont {M.}~\bibnamefont {Holland}}, \bibinfo {author}
  {\bibfnamefont {E.~A.}\ \bibnamefont {Cornell}}, \ and\ \bibinfo {author}
  {\bibfnamefont {C.}~\bibnamefont {Wieman}},\ }\href@noop {} {\bibfield
  {journal} {\bibinfo  {journal} {Physical Review Letters}\ }\textbf {\bibinfo
  {volume} {79}},\ \bibinfo {pages} {337} (\bibinfo {year} {1997})}\BibitemShut
  {NoStop}%
\bibitem [{\citenamefont {Zundel}\ \emph {et~al.}(2019)\citenamefont {Zundel},
  \citenamefont {Wilson}, \citenamefont {Malvania}, \citenamefont {Xia},
  \citenamefont {Riou},\ and\ \citenamefont {Weiss}}]{k1_k3_zundel}%
  \BibitemOpen
  \bibfield  {author} {\bibinfo {author} {\bibfnamefont {L.~A.}\ \bibnamefont
  {Zundel}}, \bibinfo {author} {\bibfnamefont {J.~M.}\ \bibnamefont {Wilson}},
  \bibinfo {author} {\bibfnamefont {N.}~\bibnamefont {Malvania}}, \bibinfo
  {author} {\bibfnamefont {L.}~\bibnamefont {Xia}}, \bibinfo {author}
  {\bibfnamefont {J.-F.}\ \bibnamefont {Riou}}, \ and\ \bibinfo {author}
  {\bibfnamefont {D.~S.}\ \bibnamefont {Weiss}},\ }\href@noop {} {\bibfield
  {journal} {\bibinfo  {journal} {Phys. Rev. Lett.}\ }\textbf {\bibinfo
  {volume} {122}},\ \bibinfo {pages} {013402} (\bibinfo {year}
  {2019})}\BibitemShut {NoStop}%
\bibitem [{\citenamefont {Roberts}\ \emph {et~al.}(2000)\citenamefont
  {Roberts}, \citenamefont {Claussen}, \citenamefont {Cornish},\ and\
  \citenamefont {Wieman}}]{k2_roberts}%
  \BibitemOpen
  \bibfield  {author} {\bibinfo {author} {\bibfnamefont {J.~L.}\ \bibnamefont
  {Roberts}}, \bibinfo {author} {\bibfnamefont {N.~R.}\ \bibnamefont
  {Claussen}}, \bibinfo {author} {\bibfnamefont {S.~L.}\ \bibnamefont
  {Cornish}}, \ and\ \bibinfo {author} {\bibfnamefont {C.~E.}\ \bibnamefont
  {Wieman}},\ }\href@noop {} {\bibfield  {journal} {\bibinfo  {journal}
  {Physical Review Letters}\ }\textbf {\bibinfo {volume} {85}},\ \bibinfo
  {pages} {728} (\bibinfo {year} {2000})}\BibitemShut {NoStop}%
\bibitem [{\citenamefont {Savage}\ \emph {et~al.}(2003)\citenamefont {Savage},
  \citenamefont {Robins},\ and\ \citenamefont {Hope}}]{k3_savage}%
  \BibitemOpen
  \bibfield  {author} {\bibinfo {author} {\bibfnamefont {C.}~\bibnamefont
  {Savage}}, \bibinfo {author} {\bibfnamefont {N.}~\bibnamefont {Robins}}, \
  and\ \bibinfo {author} {\bibfnamefont {J.}~\bibnamefont {Hope}},\ }\href@noop
  {} {\bibfield  {journal} {\bibinfo  {journal} {Physical Review A}\ }\textbf
  {\bibinfo {volume} {67}},\ \bibinfo {pages} {014304} (\bibinfo {year}
  {2003})}\BibitemShut {NoStop}%
\bibitem [{\citenamefont {Altin}\ \emph {et~al.}(2011)\citenamefont {Altin},
  \citenamefont {Dennis}, \citenamefont {McDonald}, \citenamefont {Doering},
  \citenamefont {Debs}, \citenamefont {Close}, \citenamefont {Savage},\ and\
  \citenamefont {Robins}}]{k3_altin}%
  \BibitemOpen
  \bibfield  {author} {\bibinfo {author} {\bibfnamefont {P.}~\bibnamefont
  {Altin}}, \bibinfo {author} {\bibfnamefont {G.}~\bibnamefont {Dennis}},
  \bibinfo {author} {\bibfnamefont {G.}~\bibnamefont {McDonald}}, \bibinfo
  {author} {\bibfnamefont {D.}~\bibnamefont {Doering}}, \bibinfo {author}
  {\bibfnamefont {J.}~\bibnamefont {Debs}}, \bibinfo {author} {\bibfnamefont
  {J.}~\bibnamefont {Close}}, \bibinfo {author} {\bibfnamefont
  {C.}~\bibnamefont {Savage}}, \ and\ \bibinfo {author} {\bibfnamefont
  {N.}~\bibnamefont {Robins}},\ }\href@noop {} {\bibfield  {journal} {\bibinfo
  {journal} {Physical Review A}\ }\textbf {\bibinfo {volume} {84}},\ \bibinfo
  {pages} {033632} (\bibinfo {year} {2011})}\BibitemShut {NoStop}%
\bibitem [{\citenamefont {Sinatra}\ and\ \citenamefont
  {Castin}(1998)}]{loss_sinatra1998}%
  \BibitemOpen
  \bibfield  {author} {\bibinfo {author} {\bibfnamefont {A.}~\bibnamefont
  {Sinatra}}\ and\ \bibinfo {author} {\bibfnamefont {Y.}~\bibnamefont
  {Castin}},\ }\href@noop {} {\bibfield  {journal} {\bibinfo  {journal} {The
  European Physical Journal D-Atomic, Molecular, Optical and Plasma Physics}\
  }\textbf {\bibinfo {volume} {4}},\ \bibinfo {pages} {247} (\bibinfo {year}
  {1998})}\BibitemShut {NoStop}%
\bibitem [{\citenamefont {Huang}\ and\ \citenamefont
  {Moore}(2006)}]{n-1_huang}%
  \BibitemOpen
  \bibfield  {author} {\bibinfo {author} {\bibfnamefont {Y.}~\bibnamefont
  {Huang}}\ and\ \bibinfo {author} {\bibfnamefont {M.}~\bibnamefont {Moore}},\
  }\href@noop {} {\bibfield  {journal} {\bibinfo  {journal} {Physical Review
  A}\ }\textbf {\bibinfo {volume} {73}},\ \bibinfo {pages} {023606} (\bibinfo
  {year} {2006})}\BibitemShut {NoStop}%
\bibitem [{\citenamefont {Lombardo}\ and\ \citenamefont
  {Twamley}(2015)}]{n-1_lombardo}%
  \BibitemOpen
  \bibfield  {author} {\bibinfo {author} {\bibfnamefont {D.}~\bibnamefont
  {Lombardo}}\ and\ \bibinfo {author} {\bibfnamefont {J.}~\bibnamefont
  {Twamley}},\ }\href@noop {} {\bibfield  {journal} {\bibinfo  {journal}
  {Scientific reports}\ }\textbf {\bibinfo {volume} {5}},\ \bibinfo {pages}
  {13884} (\bibinfo {year} {2015})}\BibitemShut {NoStop}%
\bibitem [{\citenamefont {Alon}\ \emph {et~al.}(2008)\citenamefont {Alon},
  \citenamefont {Streltsov},\ and\ \citenamefont
  {Cederbaum}}]{mctdhb_alon2008}%
  \BibitemOpen
  \bibfield  {author} {\bibinfo {author} {\bibfnamefont {O.~E.}\ \bibnamefont
  {Alon}}, \bibinfo {author} {\bibfnamefont {A.~I.}\ \bibnamefont {Streltsov}},
  \ and\ \bibinfo {author} {\bibfnamefont {L.~S.}\ \bibnamefont {Cederbaum}},\
  }\href@noop {} {\bibfield  {journal} {\bibinfo  {journal} {Physical Review
  A}\ }\textbf {\bibinfo {volume} {77}},\ \bibinfo {pages} {033613} (\bibinfo
  {year} {2008})}\BibitemShut {NoStop}%
\bibitem [{\citenamefont {Lode}(2016)}]{mctdhb_lode}%
  \BibitemOpen
  \bibfield  {author} {\bibinfo {author} {\bibfnamefont {A.~U.}\ \bibnamefont
  {Lode}},\ }\href@noop {} {\bibfield  {journal} {\bibinfo  {journal} {Physical
  Review A}\ }\textbf {\bibinfo {volume} {93}},\ \bibinfo {pages} {063601}
  (\bibinfo {year} {2016})}\BibitemShut {NoStop}%
\bibitem [{\citenamefont {Press}\ and\ \citenamefont {Teukolsky}(1992)}]{ark}%
  \BibitemOpen
  \bibfield  {author} {\bibinfo {author} {\bibfnamefont {W.~H.}\ \bibnamefont
  {Press}}\ and\ \bibinfo {author} {\bibfnamefont {S.~A.}\ \bibnamefont
  {Teukolsky}},\ }\href@noop {} {\bibfield  {journal} {\bibinfo  {journal}
  {Computers in Physics}\ }\textbf {\bibinfo {volume} {6}},\ \bibinfo {pages}
  {188} (\bibinfo {year} {1992})}\BibitemShut {NoStop}%
\bibitem [{\citenamefont {Press}\ \emph {et~al.}(2007)\citenamefont {Press},
  \citenamefont {Teukolsky}, \citenamefont {Vetterling},\ and\ \citenamefont
  {Flannery}}]{ark_numerical_recipes}%
  \BibitemOpen
  \bibfield  {author} {\bibinfo {author} {\bibfnamefont {W.~H.}\ \bibnamefont
  {Press}}, \bibinfo {author} {\bibfnamefont {S.~A.}\ \bibnamefont
  {Teukolsky}}, \bibinfo {author} {\bibfnamefont {W.~T.}\ \bibnamefont
  {Vetterling}}, \ and\ \bibinfo {author} {\bibfnamefont {B.~P.}\ \bibnamefont
  {Flannery}},\ }\href@noop {} {\emph {\bibinfo {title} {Numerical recipes 3rd
  edition: The art of scientific computing}}}\ (\bibinfo  {publisher}
  {Cambridge university press},\ \bibinfo {year} {2007})\BibitemShut {NoStop}%
\bibitem [{\citenamefont {Dennis}\ \emph {et~al.}(2013)\citenamefont {Dennis},
  \citenamefont {Hope},\ and\ \citenamefont {Johnsson}}]{xmds:paper}%
  \BibitemOpen
  \bibfield  {author} {\bibinfo {author} {\bibfnamefont {G.~R.}\ \bibnamefont
  {Dennis}}, \bibinfo {author} {\bibfnamefont {J.~J.}\ \bibnamefont {Hope}}, \
  and\ \bibinfo {author} {\bibfnamefont {M.~T.}\ \bibnamefont {Johnsson}},\
  }\href@noop {} {\bibfield  {journal} {\bibinfo  {journal} {Comput. Phys.
  Comm.}\ }\textbf {\bibinfo {volume} {184}},\ \bibinfo {pages} {201} (\bibinfo
  {year} {2013})}\BibitemShut {NoStop}%
\bibitem [{xmd()}]{xmds:citations}%
  \BibitemOpen
  \href@noop {} {}\bibinfo {note} {The SE and Tully's algorithm were both
  implemented in the high-level simulation language
  XMDS~\cite{xmds:paper,xmds:docu}.}\BibitemShut {Stop}%
\bibitem [{\citenamefont {Cosme}\ \emph {et~al.}(2016)\citenamefont {Cosme},
  \citenamefont {Weiss},\ and\ \citenamefont
  {Brand}}]{mctdhb_convergence_cosme}%
  \BibitemOpen
  \bibfield  {author} {\bibinfo {author} {\bibfnamefont {J.~G.}\ \bibnamefont
  {Cosme}}, \bibinfo {author} {\bibfnamefont {C.}~\bibnamefont {Weiss}}, \ and\
  \bibinfo {author} {\bibfnamefont {J.}~\bibnamefont {Brand}},\ }\href@noop {}
  {\bibfield  {journal} {\bibinfo  {journal} {Physical Review A}\ }\textbf
  {\bibinfo {volume} {94}},\ \bibinfo {pages} {043603} (\bibinfo {year}
  {2016})}\BibitemShut {NoStop}%
\bibitem [{\citenamefont {Weiss}\ \emph {et~al.}(2015)\citenamefont {Weiss},
  \citenamefont {Gardiner},\ and\ \citenamefont
  {Breuer}}]{mctdhb_convergence_weiss}%
  \BibitemOpen
  \bibfield  {author} {\bibinfo {author} {\bibfnamefont {C.}~\bibnamefont
  {Weiss}}, \bibinfo {author} {\bibfnamefont {S.~A.}\ \bibnamefont {Gardiner}},
  \ and\ \bibinfo {author} {\bibfnamefont {H.-P.}\ \bibnamefont {Breuer}},\
  }\href@noop {} {\bibfield  {journal} {\bibinfo  {journal} {Physical Review
  A}\ }\textbf {\bibinfo {volume} {91}},\ \bibinfo {pages} {063616} (\bibinfo
  {year} {2015})}\BibitemShut {NoStop}%
\bibitem [{\citenamefont {Aman}\ \emph {et~al.}(2016)\citenamefont {Aman},
  \citenamefont {DeSalvo}, \citenamefont {Dunning}, \citenamefont {Killian},
  \citenamefont {Yoshida},\ and\ \citenamefont
  {Burgd{\"o}rfer}}]{dressing_loss_aman}%
  \BibitemOpen
  \bibfield  {author} {\bibinfo {author} {\bibfnamefont {J.}~\bibnamefont
  {Aman}}, \bibinfo {author} {\bibfnamefont {B.}~\bibnamefont {DeSalvo}},
  \bibinfo {author} {\bibfnamefont {F.}~\bibnamefont {Dunning}}, \bibinfo
  {author} {\bibfnamefont {T.}~\bibnamefont {Killian}}, \bibinfo {author}
  {\bibfnamefont {S.}~\bibnamefont {Yoshida}}, \ and\ \bibinfo {author}
  {\bibfnamefont {J.}~\bibnamefont {Burgd{\"o}rfer}},\ }\href@noop {}
  {\bibfield  {journal} {\bibinfo  {journal} {Physical Review A}\ }\textbf
  {\bibinfo {volume} {93}},\ \bibinfo {pages} {043425} (\bibinfo {year}
  {2016})}\BibitemShut {NoStop}%
\bibitem [{\citenamefont {P{\l}odzie{\'n}}\ \emph {et~al.}(2017)\citenamefont
  {P{\l}odzie{\'n}}, \citenamefont {Lochead}, \citenamefont {de~Hond},
  \citenamefont {Van~Druten},\ and\ \citenamefont
  {Kokkelmans}}]{dressing_loss_plodzien}%
  \BibitemOpen
  \bibfield  {author} {\bibinfo {author} {\bibfnamefont {M.}~\bibnamefont
  {P{\l}odzie{\'n}}}, \bibinfo {author} {\bibfnamefont {G.}~\bibnamefont
  {Lochead}}, \bibinfo {author} {\bibfnamefont {J.}~\bibnamefont {de~Hond}},
  \bibinfo {author} {\bibfnamefont {N.}~\bibnamefont {Van~Druten}}, \ and\
  \bibinfo {author} {\bibfnamefont {S.}~\bibnamefont {Kokkelmans}},\
  }\href@noop {} {\bibfield  {journal} {\bibinfo  {journal} {Physical Review
  A}\ }\textbf {\bibinfo {volume} {95}},\ \bibinfo {pages} {043606} (\bibinfo
  {year} {2017})}\BibitemShut {NoStop}%
\bibitem [{\citenamefont {Bilardello}(2017)}]{heating_effect_thesis1}%
  \BibitemOpen
  \bibfield  {author} {\bibinfo {author} {\bibfnamefont {M.}~\bibnamefont
  {Bilardello}},\ }\emph {\bibinfo {title} {Heating effects and localization
  mechanism in cold Bose gases}},\ \href@noop {} {Ph.D. thesis},\ \bibinfo
  {school} {Universit{\`a} degli Studi di Trieste} (\bibinfo {year}
  {2017})\BibitemShut {NoStop}%
\bibitem [{\citenamefont {Schelle}(2009)}]{heating_effect_thesis2}%
  \BibitemOpen
  \bibfield  {author} {\bibinfo {author} {\bibfnamefont {A.}~\bibnamefont
  {Schelle}},\ }\emph {\bibinfo {title} {Environment-induced dynamics in a
  dilute Bose-Einstein condensate}},\ \href@noop {} {Ph.D. thesis},\ \bibinfo
  {school} {Universit{\'e} Pierre et Marie Curie-Paris VI} (\bibinfo {year}
  {2009})\BibitemShut {NoStop}%
\bibitem [{\citenamefont {Roberts}(2001)}]{roberts_phd}%
  \BibitemOpen
  \bibfield  {author} {\bibinfo {author} {\bibfnamefont {J.~L.}\ \bibnamefont
  {Roberts}},\ }\emph {\bibinfo {title} {Bose-Einstein condensates with tunable
  atom-atom interactions: The first experiments with 85Rb BECs}},\ \href@noop
  {} {Ph.D. thesis} (\bibinfo {year} {2001})\BibitemShut {NoStop}%
\bibitem [{\citenamefont {Jack}(2002)}]{jack}%
  \BibitemOpen
  \bibfield  {author} {\bibinfo {author} {\bibfnamefont {M.~W.}\ \bibnamefont
  {Jack}},\ }\href@noop {} {\bibfield  {journal} {\bibinfo  {journal} {Phys.
  Rev. Lett.}\ }\textbf {\bibinfo {volume} {89}},\ \bibinfo {pages} {140402}
  (\bibinfo {year} {2002})}\BibitemShut {NoStop}%
\bibitem [{\citenamefont {Chaturvedi}\ and\ \citenamefont
  {Srinivasan}(1991)}]{Chaturvedi_thermofield}%
  \BibitemOpen
  \bibfield  {author} {\bibinfo {author} {\bibfnamefont {S.}~\bibnamefont
  {Chaturvedi}}\ and\ \bibinfo {author} {\bibfnamefont {V.}~\bibnamefont
  {Srinivasan}},\ }\href@noop {} {\bibfield  {journal} {\bibinfo  {journal}
  {Phys. Rev. A}\ }\textbf {\bibinfo {volume} {43}},\ \bibinfo {pages} {4054}
  (\bibinfo {year} {1991})}\BibitemShut {NoStop}%
\bibitem [{\citenamefont {Brune}\ \emph {et~al.}(1996)\citenamefont {Brune},
  \citenamefont {Hagley}, \citenamefont {Dreyer}, \citenamefont {Maitre},
  \citenamefont {Maali}, \citenamefont {Wunderlich}, \citenamefont {Raimond},\
  and\ \citenamefont {Haroche}}]{cat_brune}%
  \BibitemOpen
  \bibfield  {author} {\bibinfo {author} {\bibfnamefont {M.}~\bibnamefont
  {Brune}}, \bibinfo {author} {\bibfnamefont {E.}~\bibnamefont {Hagley}},
  \bibinfo {author} {\bibfnamefont {J.}~\bibnamefont {Dreyer}}, \bibinfo
  {author} {\bibfnamefont {X.}~\bibnamefont {Maitre}}, \bibinfo {author}
  {\bibfnamefont {A.}~\bibnamefont {Maali}}, \bibinfo {author} {\bibfnamefont
  {C.}~\bibnamefont {Wunderlich}}, \bibinfo {author} {\bibfnamefont
  {J.}~\bibnamefont {Raimond}}, \ and\ \bibinfo {author} {\bibfnamefont
  {S.}~\bibnamefont {Haroche}},\ }\href@noop {} {\bibfield  {journal} {\bibinfo
   {journal} {Physical Review Letters}\ }\textbf {\bibinfo {volume} {77}},\
  \bibinfo {pages} {4887} (\bibinfo {year} {1996})}\BibitemShut {NoStop}%
\bibitem [{\citenamefont {Paw{\l}owski}\ \emph {et~al.}(2017)\citenamefont
  {Paw{\l}owski}, \citenamefont {Fadel}, \citenamefont {Treutlein},
  \citenamefont {Castin},\ and\ \citenamefont {Sinatra}}]{cat_pawlowski}%
  \BibitemOpen
  \bibfield  {author} {\bibinfo {author} {\bibfnamefont {K.}~\bibnamefont
  {Paw{\l}owski}}, \bibinfo {author} {\bibfnamefont {M.}~\bibnamefont {Fadel}},
  \bibinfo {author} {\bibfnamefont {P.}~\bibnamefont {Treutlein}}, \bibinfo
  {author} {\bibfnamefont {Y.}~\bibnamefont {Castin}}, \ and\ \bibinfo {author}
  {\bibfnamefont {A.}~\bibnamefont {Sinatra}},\ }\href@noop {} {\bibfield
  {journal} {\bibinfo  {journal} {Physical Review A}\ }\textbf {\bibinfo
  {volume} {95}},\ \bibinfo {pages} {063609} (\bibinfo {year}
  {2017})}\BibitemShut {NoStop}%
\bibitem [{\citenamefont {Zurek}(2003{\natexlab{b}})}]{cat_zurek}%
  \BibitemOpen
  \bibfield  {author} {\bibinfo {author} {\bibfnamefont {W.~H.}\ \bibnamefont
  {Zurek}},\ }\href@noop {} {\bibfield  {journal} {\bibinfo  {journal} {arXiv
  preprint quant-ph/0306072}\ } (\bibinfo {year}
  {2003}{\natexlab{b}})}\BibitemShut {NoStop}%
\bibitem [{\citenamefont {Louis}\ \emph {et~al.}(2001)\citenamefont {Louis},
  \citenamefont {Brydon},\ and\ \citenamefont {Savage}}]{cat_savage}%
  \BibitemOpen
  \bibfield  {author} {\bibinfo {author} {\bibfnamefont {P.}~\bibnamefont
  {Louis}}, \bibinfo {author} {\bibfnamefont {P.}~\bibnamefont {Brydon}}, \
  and\ \bibinfo {author} {\bibfnamefont {C.}~\bibnamefont {Savage}},\
  }\href@noop {} {\bibfield  {journal} {\bibinfo  {journal} {Physical Review
  A}\ }\textbf {\bibinfo {volume} {64}},\ \bibinfo {pages} {053613} (\bibinfo
  {year} {2001})}\BibitemShut {NoStop}%
\bibitem [{\citenamefont {Rammohan}\ \emph {et~al.}(2020)\citenamefont
  {Rammohan}, \citenamefont {Tiwari}, \citenamefont {Mishra}, \citenamefont
  {Kumar}, \citenamefont {Pendse}, \citenamefont {Nath}, \citenamefont
  {Eisfeld},\ and\ \citenamefont {W{\"u}ster}}]{sid_preparation}%
  \BibitemOpen
  \bibfield  {author} {\bibinfo {author} {\bibfnamefont {S.}~\bibnamefont
  {Rammohan}}, \bibinfo {author} {\bibfnamefont {S.}~\bibnamefont {Tiwari}},
  \bibinfo {author} {\bibfnamefont {A.}~\bibnamefont {Mishra}}, \bibinfo
  {author} {\bibfnamefont {A.}~\bibnamefont {Kumar}}, \bibinfo {author}
  {\bibfnamefont {A.}~\bibnamefont {Pendse}}, \bibinfo {author} {\bibfnamefont
  {R.}~\bibnamefont {Nath}}, \bibinfo {author} {\bibfnamefont {A.}~\bibnamefont
  {Eisfeld}}, \ and\ \bibinfo {author} {\bibfnamefont {S.}~\bibnamefont
  {W{\"u}ster}},\ }\href@noop {} {\bibfield  {journal} {\bibinfo  {journal} {In
  Preparation}\ } (\bibinfo {year} {2020})}\BibitemShut {NoStop}%
\bibitem [{\citenamefont {Pohl}\ \emph {et~al.}(2010)\citenamefont {Pohl},
  \citenamefont {Demler},\ and\ \citenamefont {Lukin}}]{pohl:crystal}%
  \BibitemOpen
  \bibfield  {author} {\bibinfo {author} {\bibfnamefont {T.}~\bibnamefont
  {Pohl}}, \bibinfo {author} {\bibfnamefont {E.}~\bibnamefont {Demler}}, \ and\
  \bibinfo {author} {\bibfnamefont {M.~D.}\ \bibnamefont {Lukin}},\ }\href@noop
  {} {\bibfield  {journal} {\bibinfo  {journal} {Phys. Rev. Lett.}\ }\textbf
  {\bibinfo {volume} {104}},\ \bibinfo {pages} {043002} (\bibinfo {year}
  {2010})}\BibitemShut {NoStop}%
\bibitem [{\citenamefont {van Bijnen}\ \emph {et~al.}(2011)\citenamefont {van
  Bijnen}, \citenamefont {Smit}, \citenamefont {van Leeuwen}, \citenamefont
  {Vredenbregt},\ and\ \citenamefont {Kokkelmans}}]{rickvB:adiab_crystals}%
  \BibitemOpen
  \bibfield  {author} {\bibinfo {author} {\bibfnamefont {R.~M.~W.}\
  \bibnamefont {van Bijnen}}, \bibinfo {author} {\bibfnamefont
  {S.}~\bibnamefont {Smit}}, \bibinfo {author} {\bibfnamefont {K.~A.~H.}\
  \bibnamefont {van Leeuwen}}, \bibinfo {author} {\bibfnamefont {E.~J.~D.}\
  \bibnamefont {Vredenbregt}}, \ and\ \bibinfo {author} {\bibfnamefont {S.~J.
  J. M.~F.}\ \bibnamefont {Kokkelmans}},\ }\href@noop {} {\bibfield  {journal}
  {\bibinfo  {journal} {J. Phys. B: At. Mol. Opt. Phys.}\ }\textbf {\bibinfo
  {volume} {44}},\ \bibinfo {pages} {184008} (\bibinfo {year}
  {2011})}\BibitemShut {NoStop}%
\bibitem [{\citenamefont {Schau{\ss}}\ \emph {et~al.}(2015)\citenamefont
  {Schau{\ss}}, \citenamefont {Zeiher}, \citenamefont {Fukuhara}, \citenamefont
  {Hild}, \citenamefont {Cheneau}, \citenamefont {Macr{\`\i}}, \citenamefont
  {Pohl}, \citenamefont {Bloch},\ and\ \citenamefont
  {Gro{\ss}}}]{rcrystal_schauss1}%
  \BibitemOpen
  \bibfield  {author} {\bibinfo {author} {\bibfnamefont {P.}~\bibnamefont
  {Schau{\ss}}}, \bibinfo {author} {\bibfnamefont {J.}~\bibnamefont {Zeiher}},
  \bibinfo {author} {\bibfnamefont {T.}~\bibnamefont {Fukuhara}}, \bibinfo
  {author} {\bibfnamefont {S.}~\bibnamefont {Hild}}, \bibinfo {author}
  {\bibfnamefont {M.}~\bibnamefont {Cheneau}}, \bibinfo {author} {\bibfnamefont
  {T.}~\bibnamefont {Macr{\`\i}}}, \bibinfo {author} {\bibfnamefont
  {T.}~\bibnamefont {Pohl}}, \bibinfo {author} {\bibfnamefont {I.}~\bibnamefont
  {Bloch}}, \ and\ \bibinfo {author} {\bibfnamefont {C.}~\bibnamefont
  {Gro{\ss}}},\ }\href@noop {} {\bibfield  {journal} {\bibinfo  {journal}
  {Science}\ }\textbf {\bibinfo {volume} {347}},\ \bibinfo {pages} {1455}
  (\bibinfo {year} {2015})}\BibitemShut {NoStop}%
\bibitem [{\citenamefont {Schau{\ss}}\ \emph {et~al.}(2012)\citenamefont
  {Schau{\ss}}, \citenamefont {Cheneau}, \citenamefont {Endres}, \citenamefont
  {Fukuhara}, \citenamefont {Hild}, \citenamefont {Omran}, \citenamefont
  {Pohl}, \citenamefont {Gross}, \citenamefont {Kuhr},\ and\ \citenamefont
  {Bloch}}]{rcrystal_schauss2}%
  \BibitemOpen
  \bibfield  {author} {\bibinfo {author} {\bibfnamefont {P.}~\bibnamefont
  {Schau{\ss}}}, \bibinfo {author} {\bibfnamefont {M.}~\bibnamefont {Cheneau}},
  \bibinfo {author} {\bibfnamefont {M.}~\bibnamefont {Endres}}, \bibinfo
  {author} {\bibfnamefont {T.}~\bibnamefont {Fukuhara}}, \bibinfo {author}
  {\bibfnamefont {S.}~\bibnamefont {Hild}}, \bibinfo {author} {\bibfnamefont
  {A.}~\bibnamefont {Omran}}, \bibinfo {author} {\bibfnamefont
  {T.}~\bibnamefont {Pohl}}, \bibinfo {author} {\bibfnamefont {C.}~\bibnamefont
  {Gross}}, \bibinfo {author} {\bibfnamefont {S.}~\bibnamefont {Kuhr}}, \ and\
  \bibinfo {author} {\bibfnamefont {I.}~\bibnamefont {Bloch}},\ }\href@noop {}
  {\bibfield  {journal} {\bibinfo  {journal} {Nature}\ }\textbf {\bibinfo
  {volume} {491}},\ \bibinfo {pages} {87} (\bibinfo {year} {2012})}\BibitemShut
  {NoStop}%
\bibitem [{\citenamefont {Dennis}\ \emph {et~al.}(2012)\citenamefont {Dennis},
  \citenamefont {Hope},\ and\ \citenamefont {Johnsson}}]{xmds:docu}%
  \BibitemOpen
  \bibfield  {author} {\bibinfo {author} {\bibfnamefont {G.~R.}\ \bibnamefont
  {Dennis}}, \bibinfo {author} {\bibfnamefont {J.~J.}\ \bibnamefont {Hope}}, \
  and\ \bibinfo {author} {\bibfnamefont {M.~T.}\ \bibnamefont {Johnsson}},\
  }\href@noop {} {} (\bibinfo {year} {2012}),\ \bibinfo {note}
  {http://www.xmds.org/}\BibitemShut {NoStop}%
\bibitem [{\citenamefont {Gould}\ and\ \citenamefont {Bucko}(2016)}]{c6_gould}%
  \BibitemOpen
  \bibfield  {author} {\bibinfo {author} {\bibfnamefont {T.}~\bibnamefont
  {Gould}}\ and\ \bibinfo {author} {\bibfnamefont {T.}~\bibnamefont {Bucko}},\
  }\href@noop {} {\bibfield  {journal} {\bibinfo  {journal} {J. Chem. Theory
  and Comp.}\ }\textbf {\bibinfo {volume} {12}},\ \bibinfo {pages} {3603}
  (\bibinfo {year} {2016})}\BibitemShut {NoStop}%
\end{thebibliography}%

\appendix
\section{Derivation of variational multi-orbital equations} 
\label{sec:derivation}
We here present the details of the derivation of evolution equations for the orbitals and coeffcients in the Ansatz of \eref{eq:ansatz}. The equations of motion are derived by minimization of the action given by
\begin{equation}
\begin{split}
S&=\int{dt\;d^N \mathbf{r}} \Bigg\{\Psi^*(\mathbf{r},t)\Big(\hat{H}-i\hbar\frac{\partial}{\partial t}\Big)\Psi(\mathbf{r},t)\Bigg\}\\
&-\int{dt\sum_{j=0}^{2}\lambda_{j}\Big(\int{d^N\mathbf{r} \;\Phi^{*}_{j}(\mathbf{r},t)\Phi_{j}(\mathbf{r},t)}-1 \Big)},
\end{split}
\end{equation}
where $\hat{H}$ is the Hamiltonian given by \eref{eq:hamil1} and $\lambda_{j}$ are the Lagrange multipliers. The second term on the RHS in the above equation ensures the normalization of the wavefunction at all times.

Inserting the Ansatz in \eref{eq:ansatz}, we can split the resultant expression for the action into six terms
\begin{equation}
\label{eq:actionsplit}
S=\int{dt\; d^{N}\mathbf{r} [T_{1}+T_{2}+T_{3}+T_{4}+T_{5}]}-\int{dt\; T_{6}},
\end{equation} 
which are given explicitly by
\begin{equation}
\begin{split}
&T_{1}=\sum_{i=0}^{2} |C_{i}(t)|^{2} \Phi^{*}_{i}(\mathbf{r},t) \; \hat{H}_{bec} \; \Phi_{i}(\mathbf{r},t),\\
&T_{2}=\sum_{i=1}^{2} |C_{i}(t)|^{2} |\Phi_{i}(\mathbf{r},t)|^{2} V^{(i)}_{int}({\bf{x_{i}}}-r_{j}),\\
&T_{3}=\sum_{i,j=0,i\neq j}^{2} C^{*}_{i}(t)C_{j}(t)\bra{i}\sub{\hat{H}}{ctrl}(t)\ket{j}\Phi^{*}_{i}(\mathbf{r},t)\Phi_{j}(\mathbf{r},t),\\
&T_{4}=-i\hbar\sum_{i=0}^{2}C^{*}_{i}(t)\;|\Phi_{i}(\mathbf{r},t)|^{2}\frac{\partial}{\partial t} C_{i}(t),\\
&T_{5}=-i\hbar\sum_{i=0}^{2}|C_{i}(t)|^{2}\;\Phi^{*}_{i}(\mathbf{r},t)\frac{\partial}{\partial t} \Phi_{i}(\mathbf{r},t),\\
&T_{6}=\sum_{j=0}^{2}\lambda_{j}\Big(\int{d^{N}\mathbf{r}\;\Phi^{*}_{j}(\mathbf{r},t)\Phi_{j}(\mathbf{r},t)}-1\Big),
\end{split}
\end{equation}
where the integral measure for $T_{6}$ is $\int d^{N}\mathbf{r}=\int{dr_{1}dr_{2}...dr_{N}}$, since $\Phi_{i/j}(\mathbf{r},t)$ are many body wavefunctions for $N$ Bosons.
We now invoke the structure in \eref{eq:singleorbital} for each of them, hence assuming they can be written as a product of the same single particle states for all atoms in the orbital $i$. For this, we write $\Phi_{i}(\mathbf{r},t)=\prod\limits_{j=1}^{N} \phi_{i}(r_{j},t)$. With this product ansatz and expanding $\int d^{N}\mathbf{r}$, we can simplify the above terms as discussed in the following.

Since the single particle states are normalized as stated before, so are their many-body products and the expression $|\Phi_{i}(\mathbf{r},t)|^{2}$ in $T_{4}$ is unity when integrated over all space. Many-body overlaps of the form $\int d^N\mathbf{r}\: \Phi^{*}_{i}(\mathbf{r},t) \Phi^{*}_{j}(\mathbf{r},t)$ in $T_{3}$ can be re-expressed as $N$-fold product of identical single-particle overlaps, giving $\Big[\int{d\tilde{r}}\phi^{*}_{i}(\tilde{r},t) \phi^{*}_{j}(\tilde{r},t)\Big]^{N}$. Terms $T_{1}$ and $T_{5}$ have contributions of the form $\int d^N\mathbf{r} \: \Phi^{*}_{i}(\mathbf{r},t) \; \hat{O} \; \Phi_{i}(\mathbf{r},t)$, where $\hat{O}\in\{\hat{h}_i, \frac{\partial}{\partial t} \}$, that simplify to a product of $N$ identical single particle integrals, yielding $N\;\int{d\tilde{r}}\;\phi^{*}_{i}(\tilde{r},t) \hat{\mathscr{O}} \phi_{i}(\tilde{r},t)$, where $\hat{\mathscr{O}}$ is the single particle operator acting on a single Boson. Finally, the spatial integral over $T_{2}$ contains $\int d^N\mathbf{r}\:  |\Phi_{i}(\mathbf{r},t)|^{2} V^{(i)}_{int}({\bf{x_{i}}}-\mathbf{r})$. When $\Phi_{i}(\mathbf{r},t)$ is written as product of single particle states, we get $N\int{dr}  |\phi_{i}(r,t)|^{2} V^{(i)}_{int}({\bf{x_{i}}}-r)$. Considering these factors, the action in \eref{eq:actionsplit} takes the form
\begin{equation}
\begin{split}
S&=\int{dt}\Bigg\{N\sum_{i=0}^{2}|C_{i}(t)|^{2} \int d\tilde{r} \phi^{*}_{i}(\tilde{r},t) \; \hat{h}_{i}[\phi_{i}] \; \phi_{i}(\tilde{r},t)\\
&+N\sum_{i=1}^{2} |C_{i}(t)|^{2} \int{d\tilde{r}}|\phi_{i}(\tilde{r},t)|^{2} V^{(i)}_{int}({\bf{x_{i}}}-\tilde{r})\\
&+\sum_{i,j=0,i\neq j}^{2} C^{*}_{i}(t)C_{j}(t)\bra{i}\sub{\hat{H}}{ctrl}(t)\ket{j} \Big[\int d\tilde{r} \phi^{*}_{i}(\tilde{r},t)\phi_{j}(\tilde{r},t) \Big]^{N}\\
&-i\hbar\sum_{i=0}^{2}C^{*}_{i}(t)\;\frac{\partial}{\partial t} C_{i}(t)\\
&-i\hbar N\sum_{i=0}^{2}|C_{i}(t)|^{2}\;\Big[\int{d\tilde{r}}\phi^{*}_{i}(\tilde{r},t)\frac{\partial}{\partial t} \phi_{i}(\tilde{r},t)\Big]\\
&-\sum_{j=0}^{2}\lambda_{j}\Big[\Big(\int{d\tilde{r}\;\phi^{*}_{j}(\tilde{r},t)\phi_{j}(\tilde{r},t)}\Big)^{N}-1\Big]\Bigg\},
\end{split}
\end{equation}
where the operator $\hat{h}_{i}$ is the single particle Hamiltonian for condensate atoms
\begin{equation*}
\hat{h}_{i}[\phi_{i}]=- \frac{\hbar^{2}}{2m}\nabla_{r}^{2}+ \sub{V}{ext}(r)+g(N-1)|\phi_{i}(r,t)|^{2}.
\end{equation*}
We now minimize the action above with respect to the single particle states and the coefficients using functional derivatives, via $\frac{\delta S}{\delta C^{*}_{i}}=i\hbar\frac{\partial C}{\partial t}$ and $\frac{\delta S}{\delta \phi^{*}_{i}}=i\hbar\frac{\partial \phi}{\partial t}$. From the resultant two equations, we still have to eliminate the Lagrange multiplier $\lambda_{i}$. By combining the two equations, we find that these have to satisfy
\begin{equation}
\label{eq:lambda}
\begin{split}
&\frac{\lambda_{i}}{|C_{i}(t)|^{2}}\Big[\int d\tilde{r} \phi^{*}_{i}(\tilde{r},t)\phi_{i}(\tilde{r},t) \Big]^{N-1}\\
&\hspace{0.5cm}=N^{i}_{norm}(t)\sum_{j=0,i\neq j}^{2} \frac{C_{j}(t)}{C_{i}(t)}\bra{i}\sub{\hat{H}}{ctrl}(t)\ket{j} \mathcal{M}_{ij}(t)^{N}\\
&\hspace{1cm}+N^{i}_{norm}(t)\int{d\tilde{r}\phi^{*}_{i}(\tilde{r},t) \; \hat{h}_{i} \; \phi_{i}(\tilde{r},t)}\\
&\hspace{1cm}+N^{i}_{norm}(t)\int{d\tilde{r}\phi^{*}_{i}(\tilde{r},t) V^{(i)}_{int}({\bf{x_{i}}}-\tilde{r}) \; \phi_{i}(\tilde{r},t)}\\
&\hspace{1cm}-N^{i}_{norm}(t)\int{i\hbar \; d\tilde{r}\;\phi^{*}_{i}(\tilde{r},t)\frac{\partial}{\partial t}\phi_{i}(\tilde{r},t)},
\end{split}
\end{equation}
where $N^{i}_{norm}(t)=1/\mathcal{M}_{ii}(t)$ and $\mathcal{M}_{ij}(t)=\int d\tilde{r} \; \phi^{*}_{i}(\tilde{r},t)\phi_{j}(\tilde{r},t)$. Using this expression for the Lagrange multipliers, we finally reach the evolution equations as
\begin{equation}
\label{eq:coeff_2}
\begin{split}
i\hbar\frac{\partial}{\partial t}C_{i}(t)=&\sum_{j=0,i\neq j}^{2} C_{j}(t)\bra{i}\sub{\hat{H}}{ctrl}(t)\ket{j} \mathcal{M}_{ij}(t)^{N}\\
&\hspace{-1cm}+N C_{i}(t) \int d\tilde{r} \Big[ \big(\sub{V}{ext}(\tilde{r})+V^{(i)}_{int}(\mathbf{x_{i}}-\tilde{r},t)\big)|\phi_{i}(\tilde{r},t)|^{2}\\
&\hspace{2cm}+\frac{g(N-1)}{2}|\phi_{i}(\tilde{r},t)|^{4}\Big].
\end{split}
\end{equation}
and
\begin{equation}
\label{eq:sps_3}
\begin{split}
&i\hbar \frac{\partial}{\partial t}\phi_{i}(r,t)=\\
&\hspace{0mm}\sum_{j=0,i\neq j}^{2} \Bigg\{\Bigg[\frac{C_{j}(t)}{C_{i}(t)}\bra{i}\sub{\hat{H}}{ctrl}(t)\ket{j} \Big[\int d\tilde{r} \phi^{*}_{i}(\tilde{r},t)\phi_{j}(\tilde{r},t) \Big]^{N-1}\Bigg]\\
&\hspace{10mm}\times\Bigg[\phi_{j}(r,t)-N^{i}_{norm}(t)\phi_{i}(r,t)\int d\tilde{r} \phi^{*}_{i}(\tilde{r},t)\phi_{j}(\tilde{r},t)\Bigg]\Bigg\}\\
& \\
&+\hat{h}_{i}\phi_{i}(r,t)+ V^{(i)}_{int}({\bf{x_{i}}}-r) \; \phi_{i}(r,t)\\
&-N^{i}_{norm}(t)\phi_{i}(r,t)\int{d\tilde{r}}\phi^{*}_{i}(\tilde{r},t)\Big[\sub{V}{ext}(\tilde{r})\phi_{i}(\tilde{r},t)\\
&\hspace{3cm}+g(N-1)|\phi_{i}(\tilde{r},t)|^{2}\phi_{i}(\tilde{r},t)\Big]\\
&-N^{i}_{norm}(t)\phi_{i}(r,t)\int{d\tilde{r}\phi^{*}_{i}(\tilde{r},t) V^{(i)}_{int}({\bf{x_{i}}}-\tilde{r}) \; \phi_{i}(\tilde{r},t)}.
\end{split}
\end{equation}	
We see that \eref{eq:coeff_2} implies that $\frac{\partial}{\partial t}\sum_{i=0}^{2} |C_{i}(t)|^{2}=0$ as expected and then \eref{eq:sps_3} yields $\partial N^{i}_{norm}(t)/\partial t =0$. Since we always start with normalised single particle orbitals $\int dr |\phi_i(r,t=0)|^2=1$, and these normalisations are preserved, we can just set $ N^{i}_{norm}(t)=1$ for all $i$, which we have done for \eref{eq:sps}.
 
\section{Values of parameters used for computation}
 \label{ap:numbers}
 %
We consider a bright soliton with $N=400$ atoms of $^{85}$Rb, hence the atomic mass is $m=1.419 \times 10^{-25}$ kg. Let the scattering length be tuned to $a=-5.33\times 10^{-9} $m with a Feshbach resonance. The system is made effectively 1D by tightly trapping along the radial direction with trapping frequency $\omega_{r}=300\pi \; $Hz and weak trapping along the axial direction with $\omega_{z}=100\pi \; $Hz, where we neglect the latter. The control atoms are very tightly trapped with a spread of $\sigma=0.05 \mu m$ and placed at a distance of $d=1.5 \; \mu m$ on each side of the centre of the bright soliton. 
  
 The control atoms are coupled to a Rydberg S state with principal quantum number $n_{c}=76$ using a two photon process with effective Rabi frequency $\Omega_{c}=3/(2\pi)\; $MHz . The Rydberg blockade radius under these conditions is given by $R_{b}=\sqrt[\leftroot{-2}\uproot{-2}6]{\tfrac{C_{6}}{\Omega_{c}}}\sim 4.3\mu m$ for $C_{6}\sim 3$ GHz $\times\mu m^{6}$  for the Rydberg state $76$S \cite{c6_gould}. 
 
 The atoms in the BEC soliton are dressed to a separate Rydberg S state with principal quantum number $n_{d}=55$, assuming an effective Rabi frequency $\sub{\Omega}{bec}=3/(2\pi)\; $MHz and detuning $\Delta=-500/(2\pi)\; $MHz. The Rydberg dressing causes an interaction between the soliton and the control atoms in the excited state with a potential given by $V_{int}^{(i)}(\mathbf{x_{i}}-\mathbf{r},t)$ which gives rise to \bref{eq:HI}. As mentioned there, for $i\in\{1,2\}$, $V_{int}^{(i)}(\mathbf{x_{i}}-\mathbf{r},t)$ is given by the effective potential  $\sub{U}{eff}(\mathbf{x_{i}}-\mathbf{r},t)=\alpha^{2}\Delta\Big[1-\sum\limits_{j=1}^{N}\mathcal{Q}({\bf{x_{i}}}-r_{j})/\Delta\Big]^{-1}$, where the bare van-der-Waals potential is $\mathcal{Q}({\bf{x_{i}}}-r_{j})=C_{6}/|{\bf{x_{i}}}-r_{j}|^{6}$ and the dressing parameter $\alpha=\sub{\Omega}{bec}/(2\Delta)$ \cite{sebnrick}. The position of the control atom is denoted by $\mathbf{x_{i}}$ and the position of the dressed atom by $r_{j}$. Detailed choices for all these parameters are given in \tref{tab_parameters}. 
 \begin{table*}[h]
	\centering
	\begin{tabular}{|p{0.33\linewidth}|p{0.33\linewidth}|p{0.33\linewidth}|}
		\hline
		{\bf{PARAMETER}} & {\bf{EXPRESSION}} & {\bf{VALUE/EXPRESSION}} \\
		& & \\
		\hline
		Number of Rb-85 atoms & $N$ & $400$  \\
		\hline
		Mass of Rb 85 atoms & $m$ & $1.419 \times 10^{-25}$ kg  \\
		\hline
		s-wave scattering length & $a$ & $-5.33\times 10^{-9}$m  \\
		\hline
		External trap frequencies for BEC & ($\omega_{r}$, $\omega_{z}$) & ($300\pi, 100\pi$)Hz \\
		\hline
		Spread of control atoms & $\sigma$ & $0.05 \; \mu m$ \\
		\hline
		Distance between control atoms & $2d$ & $3 \; \mu m$ \\
		\hline
		Rydberg state to which the control atoms are coupled& $n_{c}$& $76\; S$ \\
		\hline
		Rabi frequency for control atom excitation& $\Omega_{c}$ & $(3/\sqrt{2}\pi)\times 10^{6}$ Hz\\
		\hline
		Blockade radius for control atoms& $R_{b}=\sqrt[\leftroot{-2}\uproot{-2}6]{\tfrac{C_{6}}{\Omega_{c}}}\sim 4.3\mu m$& $4.3 \mu m$\\
		\hline
		Rydberg state to which BEC atoms are dressed & $n_{d}$ & $55 \; S$\\
		\hline
		Rabi frequency for BEC atom dressing &$\Omega_{bec}$ & $3/(2\pi)\;$ MHz \\
		\hline
		Detuning for dressing excitation & $\Delta$ & $-500/(2\pi)\;$ MHz\\
		\hline 
		Dressing parameter & $\alpha=\Omega_{bec}/(2\Delta)$ & $-3\times 10^{-3}$\\
		\hline 
		$C_{6}$ coefficient for interaction between two atoms in a $76$S Rydberg state & $C_{6}$ & $3$ GHz$\times\mu m^{6}$\\
		\hline
		Effective interaction potential between control atom ($i$) and Rydberg dressed BEC atoms ($j$) & $\sub{U}{eff}(\mathbf{x_{i}}-\mathbf{r},t)$ & $\alpha^{2}\Delta\Big[1-\sum\limits_{j=1}^{N}\frac{C_{6}}{(|{\bf{x_{i}}}-r_{j}|^{6})\Delta}\Big]^{-1}$ \\
		\hline
	\end{tabular}
	\caption{A tabular representation for the parameters and their values or expressions used in the simulation.
	\label{tab_parameters}
	}
\end{table*}
%
\section{Calculation of effective loss rates} 
\label{ap:lossrates}

Using \eref{eq:lme1}, we can find the master equation for elements of the density matrix using the Ansatz given in \eref{eq:lme2} and inserting the Ansatz for single particle states for particles in orbitals $\Phi_{L}(\mathbf{r},t)$ and $\Phi_{R}(\mathbf{r},t)$ respectively as $\phi_{L}(r,t)=(1/\sqrt{2\xi_{0}}) \;\sech{[(r+r_{0})/\xi_{0}]}$ and $\phi_{R}(r,t)=(1/\sqrt{2\xi_{0}}) \;\sech{[(r-r_{0})/\xi_{0}]}$ 
since we are considering the bright soliton solution of the BEC. This would give us the equation
\begin{equation}
\begin{split}
\frac{\partial \rho_{kl;mn}}{\partial t}&= \mathcal{T}_{0;klmn} \; \rho_{kl;mn} \\
&\hspace{-1.5cm}+\bar{\kappa}_{1}\mathcal{T}_{1;km} \; \rho_{(k+1)l;(m+1)n}+\bar{\kappa}_{1}\mathcal{T}_{1;ln} \; \rho_{k(l+1);m(n+1)}\\
&\hspace{-1.5cm}+\bar{\kappa}_{2,L}\;\mathcal{T}_{2;km} \; \rho_{(k+2)l;(m+2)n}+\bar{\kappa}_{2,R}\;\mathcal{T}_{2;ln} \; \rho_{k(l+2);m(n+2)}\\
&\hspace{-1.5cm}+\bar{\kappa}_{3,L}\;\mathcal{T}_{3;km} \; \rho_{(k+3)l;(m+3)n}+\bar{\kappa}_{3,R}\;\mathcal{T}_{3;ln} \; \rho_{k(l+3);m(n+3)}.
\end{split}
\label{eq:decoherence_appendix}
\end{equation}
where the density matrix elements $\rho_{kl;mn}$ are time dependent. The coefficients in the equation above are given as follows in Eqs.~(\ref{eq:rate_t0}-\ref{eq:loss_coeff_redef}).
The coefficients involved in diagonal terms are
\begin{widetext}
\begin{equation}
\begin{split}
\mathcal{T}_{0;klmn}&=\Big[\frac{1}{i\hbar}\Big(\bar{E}_{L}(k-m)+\bar{E}_{R}(l-n)+\frac{g}{2}\overline{\rho_{L}^{2}}\{k(k-1)-m(m-1)\}+\frac{g}{2}\overline{\rho_{R}^{2}}\{l(l-1)-n(n-1)\}\Big)\\
&\hspace{0cm}-\frac{1}{2}\kappa_{1}(k+l+m+n)\\
&\hspace{0cm}-\frac{1}{2}\kappa_{2}\Big(k(k-1)\overline{\rho_{L}^{2}}+m(m-1)\overline{\rho_{L}^{2}}+l(l-1)\overline{\rho_{R}^{2}}+n(n-1)\overline{\rho_{R}^{2}}\Big)\\
&\hspace{0cm}-\frac{1}{2}\kappa_{3}\Big(k(k-1)(k-2)\overline{\rho_{L}^{3}}+m(m-1)(m-2)\overline{\rho_{L}^{3}}+l(l-1)(l-2)\overline{\rho_{R}^{3}}+n(n-1)(n-2)\overline{\rho_{R}^{3}}\Big)\Big],
\end{split}
\label{eq:rate_t0}
\end{equation}
\end{widetext}
where  $\bar{E}_{L/R}$, $\overline{\rho_{L/R}^{2}}$ and $\overline{\rho_{L/R}^{3}}$ are defined in \eref{eq:nlr}.
Note that for the particular problem that we look at over here, $\bar{E}_{L}=\bar{E}_{R}$, $\overline{\rho_{L}^{2}}=\overline{\rho_{R}^{2}}$ and $\overline{\rho_{L}^{3}}=\overline{\rho_{R}^{3}}$. However, we have kept these terms distinct in the expression above to demonstrate the origin of terms.
Next, the terms which connect the coefficients int the density matrix to coefficients separated by loss of one Boson are
\begin{equation}
\begin{split}
&\mathcal{T}_{1;km}=N\sqrt{(k+1)(m+1)},\\
&\mathcal{T}_{1,ln}=N\sqrt{(l+1)(n+1)}.
\end{split}
\label{eq:rate_t1}
\end{equation}
Similarly, the terms which connect the coefficients of the density matrix to coefficients separated by the loss of two Bosons are
\begin{equation}
\begin{split}
&\mathcal{T}_{2;km}=N^{2} \; \sqrt{(k+1)(k+2)(m+1)(m+2)},\\
&\mathcal{T}_{2,ln}=N^{2} \; \sqrt{(l+1)(l+2)(n+1)(n+2)}.
\end{split}
\end{equation}
and lastly the terms which connect the coefficients of the density matrix to coefficients separated by the loss of three Boson are
\begin{equation}
\begin{split}
&\mathcal{T}_{3;km}=N^{3} \; \sqrt{(k+1)(k+2)(k+3)(m+1)(m+2)(m+3)},\\
&\mathcal{T}_{3;ln}=N^{3} \; \sqrt{(l+1)(l+2)(l+3)(n+1)(n+2)(n+3)}.
\end{split}
\end{equation}
In the above expressions, the coefficients $N_{L/R}$ are defined, due to the symmetry of the system at hand, as
\begin{equation}
\begin{split}
&{\bar{E}_{L}}=\int{dr \; \phi^{*}_{L}(r) \; \hat{h}_{f} \; \psi_{L}(r)}=\int{dr \; \phi^{*}_{R}(r) \; \hat{h}_{f} \; \phi_{R}(r)}={\bar{E}_R}\\
&{\overline{\rho_{L}^{2}}}=\int{dr \; |\phi_{L}(r)|^{4}}=\int{dr \; |\phi_{R}(r)|^{4}}={\overline{\rho_{R}^{2}}}\\
&{\overline{\rho_{L}^{3}}}=\int{dr \; |\phi_{L}(r)|^{6}}=\int{dr \; |\phi_{R}(r)|^{6}}={\overline{\rho_{R}^{3}}},
\end{split}
\label{eq:nlr}
\end{equation}
where $\hat{h}_{f}=-(\hbar^{2}/2m)\nabla^{2}_{r}+V_{ext}(r)$ in the above expression is the single particle Hamiltonian for non-interacting Bosons
{and the notation ${\overline{\rho^2}}$ indicates a squared density averaged over the soliton mode.}
Finally, the redefined loss coefficients used in \eref{eq:decoherence} and \eref{eq:decoherence_appendix} are given by
\begin{equation}
\begin{split}
&\bar{\kappa}_{1}=\kappa_{1},\\
&\bar{\kappa}_{2}=\bar{\kappa}_{2,L/R}=\overline{\rho_{L/R}^{2}} \; \kappa_{2},\\
&\bar{\kappa}_{3}=\bar{\kappa}_{3,L/R}=\overline{\rho_{L/R}^{3}}\; \kappa_{3}.
\end{split}
\label{eq:loss_coeff_redef}
\end{equation}
\end{document}